\documentclass{aa}  

\usepackage{graphicx}
\usepackage{txfonts}
\begin{document}

%%%%%%%%%%%%%%%%%%%%%%%%%%%%%%%%%%%%%%%%%%% 
%                                                       HEADER
%%%%%%%%%%%%%%%%%%%%%%%%%%%%%%%%%%%%%%%%%%% 

%TITLE
\title{Evolution of the T~Tauri star population in the Lupus association}
\subtitle{}

%AUTHORS
\author{
P. A. B. Galli \inst{1}
\and
C. Bertout\inst{2}
\and 
R. Teixeira\inst{1}
\and
C. Ducourant\inst{3}
}

%INSTITUTE
\institute{Instituto de Astronomia, Geof\'isica e Ci\^encias Atmosf\'ericas, Universidade de S\~ao Paulo, Rua do Mat\~ao, 1226 - Cidade Universit\'aria, 05508-900, S\~ao Paulo - SP, Brazil.\\
\email{galli@astro.iag.usp.br}
\and
Institut d'Astrophysique, 98bis, Bd. Arago, 75014 Paris, France.
\and
Observatoire Aquitain des Sciences de l 'Univers, CNRS-UMR 5804, BP 89, Floirac, France.
}

\date{}

%%%%%%%%%%%%%%%%%%%%%%%%%%%%%%%%%%%%%%%%%%% 
%                                                               ABSTRACT
%%%%%%%%%%%%%%%%%%%%%%%%%%%%%%%%%%%%%%%%%%% 
 
\abstract
%Context
{}
%Aims
{In a recent study, we derived individual distances for 109 pre-main sequence stars that define the Lupus kinematic association of young stars. Here, we use these new distances to derive the masses and ages of Lupus T~Tauri stars with the aim of better constraining the lifetime of their circumstellar disks.}
%Methods
{Using the photometric and spectroscopic information available in the literature, we computed the photospheric luminosity of 92 T~Tauri stars in the Lupus association. Then, we estimated their masses and ages from theoretical evolutionary models. Based on Monte Carlo simulations and statistical tests, we compare the mass and age distribution of the classical T~Tauri stars (CTTS) and weak-line T~Tauri (WTTS) in our sample.}
%Results
{We show that the CTTSs are on average younger than the WTTSs and that the probability that both T~Tauri subclasses are drawn from the same mass and age parental distribution is very low. Our results favor the scenario proposed earlier for the Taurus-Auriga association, where the CTTSs evolve into WTTSs when their disks are fully accreted by the star. Based on an empirical disk model, we find that the average disk lifetime for the T~Tauri stars in the Lupus association is  $\tau_{d}=3\times10^{6}\,(M_*/M_{\odot})^{0.55}$~yr. }
%Conclusions
{We find evidence that the average lifetime of the circumstellar disks in the Lupus association is shorter than in the Taurus-Auriga association and discuss the implications of this result.  }

%Keywords

\maketitle

%%%%%%%%%%%%%%%%%%%%%%%%%%%%%%%%%%%%%%%%%%% 
%                                                       SECTION 1
%%%%%%%%%%%%%%%%%%%%%%%%%%%%%%%%%%%%%%%%%%% 
\section{Introduction}

T~Tauri stars (TTSs) are late-type low-mass ($M\lesssim2M_{\odot}$) pre-main sequence stars that are usually associated with molecular clouds. They tend to be found in young associations \citep{Joy(1945),Herbig(1962),Bertout(1989)}. They are classified into one of two types: (i) the classical T~Tauri stars (CTTSs) that show strong emission lines indicative of accretion from a circumstellar disk, and (ii) the weak-line T~Tauri stars (WTTSs) that exhibit enhanced magnetic activity and show no signs of accretion. To explain the existence of both subgroups, it has been hypothesized for a long time \citep{Walter(1988), Bertout(1989)} that  the CTTSs become WTTSs when their disks dissipate. Although this is not a problem for the older population ($t>10$~Myr) of WTTSs, this picture implies that the timescale of disk dissipation should be a few Myr to justify the CTTSs and WTTSs coexisting \citep[see, e.g., Fig.~15 of][]{Kenyon(1995)} in the Hertzsprung-Russell diagram (HRD). Indeed, previous studies \citep{Haisch(2001),Takagi(2014)} indicate that the disk lifetime in different clusters is typically in the range 3-6~Myr. However, \citet{Fedele(2010)} demonstrate in a more detailed study that mass accretion and dust dissipation in protoplanetary disks occur on a typical timescale of 2.3-3.0~Myr (see Fig.~4 of their paper), which implies a rapid disk evolution.

Some progress in determining the lifetime of circumstellar disks of TTSs was achieved by \citet{Bertout(2007)}, who used the kinematic parallaxes derived from a kinematic study of the Taurus-Auriga T association \citep{Bertout(2006)} and calculated the stellar luminosities, masses, and ages to readdress the question of the relationship between CTTSs and WTTSs. As a result, they obtained, for the first time, the lifetime of a circumstellar disk in terms of the mass of the parent star, $\tau_{d}=4\times10^{6}\,(M/M_{\odot})^{0.75}$~yr, and demonstrated that the observed mass and age distributions of the TTSs in the Taurus-Auriga association can be explained by assuming that the CTTSs evolve into WTTSs when their disks are fully accreted by the star. 

While supporting the current scenario of disk evolution for the Taurus-Auriga T association, this first result needs to be confirmed and compared to other star-forming regions (SFRs). One reason for extending this study to other SFRs is that the origin of the initial mass function remains unclear, but is expected to depend on the intrinsic properties of each SFR, such as clustering, mass segregation, and binarity \citep{Bonnell(2007)}. Moreover,  the evolution of circumstellar (protoplanetary) disks depends on environmental criteria \citep{Rosotti(2014)}, and a comparison with other SFRs could provide important clues for constraining the timescale of planet formation in different environments. For example, \citet{Marinas(2013)} conclude that the dependence of disk fraction on stellar mass in Taurus and the NGC~2264 open cluster is different (see Fig.~10 of their paper), and they argue that the lower disk fractions in NGC~2264 might be due to the much higher stellar density than in Taurus. 

In the following, we perform a study similar to the one by \citet{Bertout(2006)} for the TTSs in the Lupus SFR. The Lupus complex of molecular clouds is located in an environment surrounded by the more massive stars from the nearby Scorpius-Centaurus OB association \citep{deZeeuw(1999)}. This makes the Lupus SFR an interesting target for comparative studies with Taurus-Auriga, because of the different environmental and dynamical effects (e.g., solar winds and UV radiation) caused by the existence of hot stars in this region. In a recent work, \citet[][hereafter Paper~I]{Galli(2013)} investigated the kinematics of the Lupus SFR and performed a membership analysis. They identified 109 pre-main sequence stars that define the comoving association of the Lupus SFR (hereafter, the Lupus association) based on a new convergent point search method \citep{Galli(2012)} and a modified version of the k-NN method. Then, they used stellar proper motions and radial velocities to calculate individual distances for all group members. In this paper, we use these newly derived kinematic distances to accurately determine the masses and ages of the TTSs in the Lupus association and constrain the lifetime of their circumstellar disks.

%%%%%%%%%%%%%%%%%%%%%%%%%%%%%%%%%%%%%%%%%%% 
%                                                       SECTION 2
%%%%%%%%%%%%%%%%%%%%%%%%%%%%%%%%%%%%%%%%%%% 
\section{Physical properties of the comoving stars}

The sample of stars used in this study is based on the 109 comoving stars of the Lupus association identified in Paper~I.  One star, SSTc2dJ154013.7-340142, which was included in our sample as a YSO candidate member of the Lupus clouds \citep[see][]{Comeron(2009)}, is excluded from the upcoming analysis, because it turned out to be a background giant star \citep[see][]{Comeron(2013)}. Our sample consists of  38 CTTSs (or CTTSs systems), 67 WTTSs (or WTTS systems), and 3 Herbig Ae/Be stars (HAeBe). These numbers correct those given in Sect.~7.1 of Paper I, where some objects were either misidentified or have been reclassified in the meantime. The results and conclusions given in Paper I are obviously not affected by this minor correction. We restricted our analysis in subsequent sections to the TTSs in the Lupus association (i.e., CTTSs and WTTSs) and excluded the HAeBes in our sample that appear in small numbers and are not statistically significant for this study.

The TTS subclass (CTTS or WTTS) from the stars in our sample were taken from the literature (see Table~1 for references), and no attempt has been made to reclassify them. However, we compiled the H$\alpha$ equivalent widths (EW) of our targets and calculated the near-infrared excess emission that is indicative of optically thick accretion disks to investigate whether the given TTS subclass was consistent with other diagnostics. We measure the excess emission by
\begin{equation}
 \Delta(H-K)=(H-K)-(H-K)_{0}\, , 
\end{equation}
where $(H-K)$ is the dereddened color, and $(H-K)_{0}$  the intrinsic color of the star \citep[see][]{Hughes(1994),Wichmann(1997a)}. 

We used the standard limit of 10{\AA} \citep[see, e.g.,][]{Appenzeller(1989)} to distinguish between CTTSs ($EW(H\alpha)\geq10{\AA}$) and WTTSs ($EW(H\alpha)<10{\AA}$), together with the limit of $\Delta(H-K)\geq0.09$ found by \citet{Edwards(1993)} for a sample of TTSs that show spectroscopic evidence of accretion. Figure~1 \textit{(upper panel)} illustrates our analysis and clearly separates the accreting TTSs (i.e., CTTSs) from the non-accreting TTSs (i.e., WTTSs). Alternatively, \citet{White(2003)} use the $H\alpha$ EW with a different cutoff that depends on the spectral type to distinguish between CTTSs and WTTSs. They propose that a TTS is indeed a CTTS if $EW(H\alpha)\geq3{\AA}$ for K0-K5 stars, $EW(H\alpha)\geq10{\AA}$ for K7-M2.5 stars, $EW(H\alpha)\geq20{\AA}$ for M3-M5.5 stars, and $EW(H\alpha)\geq40{\AA}$ for M6-M7.5 stars. We applied this classification scheme to the stars with spectral type later than K0 in our sample as shown in Figure~1 \textit{(lower panel)}. We modified the first spectral type interval slightly to include two K6 stars in our sample and provide a TTS subclass to all stars. Our analysis presented in Fig.~1 not only confirms the TTS subclass found in several sources in the literature, but also gives us more confidence when discussing the different properties of these two subgroups in the rest of this paper. Only one star (Sz~76) would be misidentified by this alternative classification (see Fig.~1), because its $EW(H\alpha)=10.3{\AA}$ \citep[see][]{Hughes(1994)} is marginally above the cutoff limit for its spectral type. However, Sz~76 has been classified recently as a WTTS by \citet{Wahhaj(2010)} and a Class~III source by \citet{Marti(2011)}. Moreover, its excess emission $\Delta(H-K)$ is clearly beyond the cutoff limit of 0.09 between accreting and non-accreting TTSs \citep[see][]{Edwards(1993)}. We therefore retain Sz~76 as a WTTSs in our study.

%%%%%%%%%%%%%%%%%%%%%%%%%%%%%%%%%%%%%%%%%%% 
%                                                       FIGURE 1
%%%%%%%%%%%%%%%%%%%%%%%%%%%%%%%%%%%%%%%%%%% 
\begin{figure}[!hbtp]
\begin{center}
\includegraphics[width=0.50\textwidth]{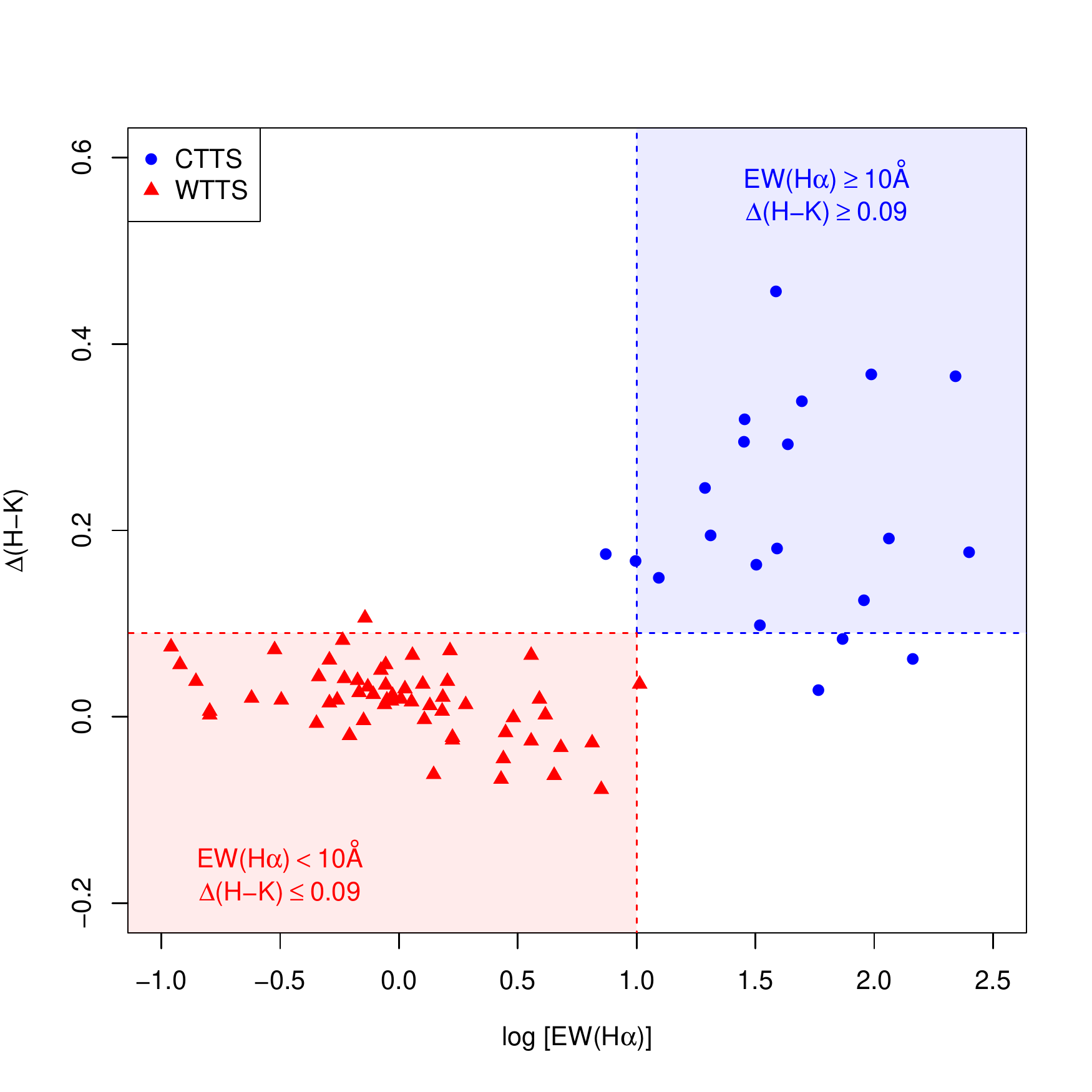}
\includegraphics[width=0.50\textwidth]{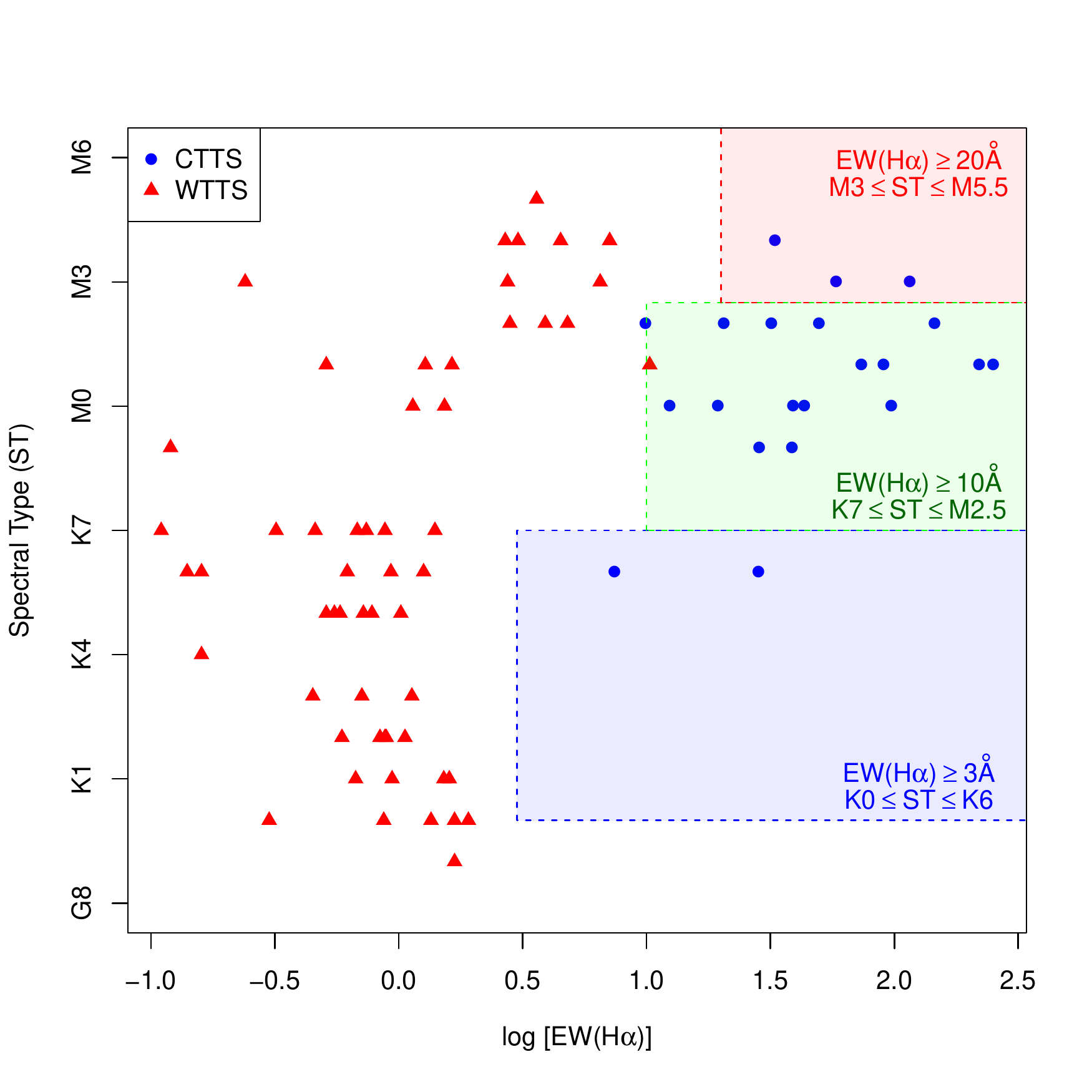}
\caption{Plots of the H$\alpha$ EWs against the infrared excess emission $\Delta (H-K)$ (\textit{upper panel}) and spectral type \textit{(lower panel)} to distinguish between CTTSs and WTTSs.
\label{Halpha}}
\end{center}
\end{figure}

The first step toward accurately computing the photospheric luminosity of stars is to derive their visual extinction. We computed the stellar extinction $A_{V}$ from the color excess in $(V-I_{C})$, $(V-R_{C}),$ and $(J-K)$ for the WTTSs in our sample, following the procedure described in Sect.~4.2 of \citet{Kenyon(1995)}. We set the extinction to zero when the derived $A_{V}$ estimates yield a non-physical negative value or a non-significant result (i.e., $A_{V}<\sigma_{A_{V}}$). We took the mean of all non-zero $A_{V}$ estimates as the final result with the standard deviation as the formal uncertainty. We used a total-to-selective extinction ratio $R_{V}$ equal to $3.1$ \citep{Savage(1979)} and adopted the reddening relations for the various colors from \citet{Bessell(1988)}. The extinction for the CTTSs in our sample is estimated from the $(R_{C}-I_{C})$ color excess with $A_{V}=4.76\cdot[(R_{C}-I_{C})-(R_{C}-I_{C})_{0}]$ as described in Sect.~2 of \citet{Cieza(2005)}. 

To gain more confidence in our results, we decided to compute our $A_{V}$ estimates again by using the same dereddening method for CTTSs and WTTSs. In this alternative approach we used only the optical color excess in $(V-I_{C})$, $(V-R_{C}),$ and $(R_{C}-I_{C}$) for all stars in the sample (i.e., CTTSs and WTTSs). We concluded that both strategies yield compatible results and that our choice of using different methods to deredden CTTSs and WTTSs (as described above) has no significant impact later on the determination of stellar masses and ages.

We used the individual parallaxes given in Tables~6 and 7 of Paper~I to derive the photospheric luminosity of the Lupus comoving stars. The stellar luminosities for the CTTSs in our sample were calculated from the $I_{C}$ band, because the standard procedure of using the $J$ flux systematically overestimates their luminosities \citep[see discussion in][]{Cieza(2005)}. For the WTTSs, we derived the stellar luminosities from the $V$ flux when the $I_{C}$ flux was not available. We verified that both procedures yield equivalent results within 5\% when both measurements are available.  We used the intrinsic colors, bolometric corrections, and temperatures for each spectral type given by \citet{Kenyon(1995)}. The uncertainties on the derived luminosities take the various sources of photometric errors
into account, but they are strongly dominated by the error budget of parallaxes. No optical photometry is available for two of our WTTSs (RXJ1546.7-3459 and RXJ1612.3-4012). We thus estimate their luminosity from the $J$ flux given by the 2MASS catalog \citep{Cutri(2003)}. Another star, Sz~102, was removed from the sample because it is believed to be associated with bipolar jets and an edge-on disk that makes our $A_{V}$ estimate very imprecise \citep{Krautter(1986),Hughes(1994)}. Moreover, its spectral type is not well-defined, and our analysis underestimates the luminosity, putting it below the zero-age main sequence (ZAMS). 

To better compare our results with those derived for the Taurus-Auriga association, we used the same completeness limits as \citet{Bertout(2007)} by removing the stars (or stellar components) that have derived luminosities lower than $0.15L_{\odot}$ and spectral types later than M4. These values correspond to $V\simeq16$~mag and $T_{eff}\simeq3300$~K. The final sample of stars with known photometry and spectral type that fulfill the above-mentioned criteria contains 30 CTTSs and 62 WTTSs. Physical properties of the Lupus stars discussed in this section are given in Table~1. Eighteen stars in our sample were identified as binary systems in the literature. The photometry of these stars was corrected for binarity with the information (mainly the flux ratio of the components) provided by \citet{Reipurth(1993)}, \cite{Brandner(1996)}, \citet{Ghez(1997)}, \citet{Guenther(2001)}, \citet{Torres(2006)}, and \citet{Romero(2012)}.

%%%%%%%%%%%%%%%%%%%%%%%%%%%%%%%%%%%%%%%%%%% 
%                                                       TABLE 1
%%%%%%%%%%%%%%%%%%%%%%%%%%%%%%%%%%%%%%%%%%% 

%This line avoids an unexpected page break in the table below
\begin{longtab}
\end{longtab}

\setlength\LTleft{-15pt}            % default: \fill
\setlength\LTright{-15pt} 

\begin{longtab}
\renewcommand\thetable{1} 
\tiny{
\begin{longtable}{lccccccccccccccc}

\caption{Physical properties of the Lupus comoving stars.
\label{tab1}  }\\
\hline
Star&ST&$T_{eff}$&$EW(H\alpha)$&$A_{V}$&$L/L_{\odot}$&Filter&$M/M_{\odot}$&$\log t$&T~Tauri&$\log\tau$&Ref.\\
&&(K)&({\AA})&(mag)&&&&(t in Myr)&Subclass&($\tau$ in Myr)&\\
\hline
\endfirsthead
\caption{continued.}\\
\hline
Star&ST&$T_{eff}$&$EW(H\alpha)$&$A_{V}$&$L/L_{\odot}$&Filter&$M/M_{\odot}$&$\log t$&T~Tauri&$\log\tau$&Ref.\\
&&(K)&({\AA})&(mag)&&&&(t in Myr)&Subclass&($\tau$ in Myr)&\\
\hline
\endhead
\hline
\endfoot

RXJ1508.8-3715  &       K5      &       4350    &       1.02    &$      0.25    \pm     0.17    $&$     0.23    ^{+     0.16    }_{     -0.09   }$&     V       &$      0.80    \pm     0.15    $&$     7.59    \pm     0.26    $&      WTTS    &$      6.48    \pm     0.14    $&      1,2     \\
RXJ1518.4-3738  &       K0      &       5250    &       1.91    &$      0.32    \pm     0.10    $&$     0.75    ^{+     0.54    }_{     -0.28   }$&     Ic      &$      0.92    \pm     0.18    $&$     7.52    \pm     0.19    $&      WTTS    &$      6.51    \pm     0.15    $&      1,2,3   \\
RXJ1524.5-3652  &       K0      &       5250    &       1.35    &$      0.44    \pm     0.07    $&$     1.24    ^{+     1.00    }_{     -0.47   }$&     Ic      &$      1.17    \pm     0.27    $&$     7.29    \pm     0.32    $&      WTTS    &$      6.57    \pm     0.17    $&      1,2,3   \\
RXJ1525.0-3604  &       K1      &       5080    &       0.94    &$      0.53    \pm     0.18    $&$     2.70    ^{+     2.89    }_{     -1.22   }$&     Ic      &$      1.68    \pm     0.45    $&$     6.77    \pm     0.39    $&      WTTS    &$      6.65    \pm     0.20    $&      1,2,3   \\
RXJ1525.5-3613  &       K2      &       4900    &       0.59    &$      0.51    \pm     0.21    $&$     1.17    ^{+     1.14    }_{     -0.52   }$&     Ic      &$      1.29    \pm     0.34    $&$     7.06    \pm     0.42    $&      WTTS    &$      6.59    \pm     0.20    $&      1,2,4   \\
RXJ1531.3-3329  &       G9      &       5410    &       1.68    &$      1.54    \pm     0.04    $&$     4.30    ^{+     4.91    }_{     -1.86   }$&     Ic      &$      1.73    \pm     0.54    $&$     6.83    \pm     0.34    $&      WTTS    &$      6.66    \pm     0.23    $&      1,2,3   \\
RXJ1534.6-4003  &       K5      &       4350    &       0.78    &$      0.33    \pm     0.16    $&$     0.25    ^{+     0.32    }_{     -0.12   }$&     Ic      &$      0.82    \pm     0.16    $&$     7.50    \pm     0.36    $&      WTTS    &$      6.48    \pm     0.15    $&      1,2,3   \\
RXJ1540.7-3756  &       K6      &       4205    &       0.93    &0.00:&$        0.36    ^{+     0.43    }_{     -0.16   }$&     Ic      &$      0.89    \pm     0.15    $&$     7.05    \pm     0.58    $&      WTTS    &$      6.50    \pm     0.13    $&      1,2,5   \\
RXJ1544.5-3521  &       K5      &       4350    &       0.51    &$      0.59    \pm     0.19    $&$     0.31    ^{+     0.42    }_{     -0.15   }$&     V       &$      0.90    \pm     0.25    $&$     7.36    \pm     0.39    $&      WTTS    &$      6.50    \pm     0.21    $&      1,2     \\
Sz73    &       M0      &       3850    &       97.20   &$      2.71    \pm     0.11    $&$     0.54    ^{+     22.54   }_{     -0.39   }$&     Ic      &$      0.57    \pm     0.08    $&$     6.34    \pm     0.58    $&      CTTS    &$      6.40    \pm     0.11    $&      2,6     \\
GQLup*  &       K7-M0   &       3955    &       38.60   &0.00:&$        2.51    ^{+     3.96    }_{     -1.20   }$&     Ic      &$      0.73    \pm     0.13    $&$     5.82    \pm     0.35    $&      CTTS    &$      6.45    \pm     0.13    $&      2,6,7   \\
RXJ1549.9-3629  &       K2      &       4900    &       1.06    &$      0.28    \pm     0.18    $&$     1.92    ^{+     3.41    }_{     -1.03   }$&     Ic      &$      1.55    \pm     0.55    $&$     6.80    \pm     0.52    $&      WTTS    &$      6.63    \pm     0.26    $&      1,2,4   \\
RXJ1552.3-3819  &       K7      &       4060    &       0.68    &0.00:&$        0.39    ^{+     0.42    }_{     -0.16   }$&     Ic      &$      0.80    \pm     0.12    $&$     6.87    \pm     0.54    $&      WTTS    &$      6.48    \pm     0.12    $&      1,2,4   \\
RXJ1605.7-3905  &       K0      &       5250    &       0.30    &$      0.15    \pm     0.07    $&$     1.49    ^{+     2.63    }_{     -0.75   }$&     Ic      &$      1.25    \pm     0.46    $&$     7.20    \pm     0.46    $&      WTTS    &$      6.58    \pm     0.27    $&      2,3     \\
F304    &       K6      &       4205    &       1.26    &0.00:&$        0.23    ^{+     0.19    }_{     -0.09   }$&     Ic      &$      0.79    \pm     0.15    $&$     7.42    \pm     0.36    $&      WTTS    &$      6.47    \pm     0.14    $&      1,2,4   \\
RXJ1608.5-3847  &       M2      &       3580    &       5.98    &$      1.00    \pm     0.07    $&$     0.72    ^{+     2.20    }_{     -0.42   }$&     Ic      &$      0.40    \pm     0.05    $&$     6.04    \pm     0.23    $&      CTTS/TD &$      6.31    \pm     0.11    $&      1,2,4,8 \\
RXJ1610.0-4016  &       K2      &       4900    &       0.88    &$      0.52    \pm     0.28    $&$     2.05    ^{+     3.41    }_{     -1.12   }$&     Ic      &$      1.59    \pm     0.57    $&$     6.76    \pm     0.51    $&      WTTS    &$      6.64    \pm     0.26    $&      1,2,3   \\
SZ121   &       M3      &       3470    &       6.50    &$      0.98    \pm     0.18    $&$     0.52    ^{+     10.39   }_{     -0.38   }$&     Ic      &$      0.31    \pm     0.08    $&$     6.17    \pm     0.31    $&      WTTS    &$      6.25    \pm     0.20    $&      2,6     \\
RXJ1613.0-4004  &       K7      &       4060    &       0.74    &$      1.34    \pm     0.04    $&$     0.26    ^{+     0.58    }_{     -0.14   }$&     Ic      &$      0.80    \pm     0.13    $&$     7.15    \pm     0.71    $&      WTTS    &$      6.48    \pm     0.13    $&      1,2,4   \\
RXJ1511.0-3252AB        &       K6      &       4205    &       0.62    &0.00:&$        0.83    ^{+     0.64    }_{     -0.30   }$&     Ic      &$      0.94    \pm     0.12    $&$     6.53    \pm     0.43    $&      WTTS    &$      6.51    \pm     0.11    $&      1,2,4   \\
RXJ1511.6-3550  &       K5      &       4350    &       0.58    &0.00:&$        0.39    ^{+     0.27    }_{     -0.13   }$&     Ic      &$      0.96    \pm     0.15    $&$     7.23    \pm     0.40    $&      WTTS    &$      6.52    \pm     0.12    $&      1,2,4   \\
RXJ1512.6-3417  &       K5      &       4350    &       ---     &$      0.56    \pm     0.19    $&$     0.31    ^{+     0.31    }_{     -0.14   }$&     V       &$      0.90    \pm     0.19    $&$     7.36    \pm     0.44    $&      WTTS    &$      6.50    \pm     0.16    $&      2,9     \\
HD135127        &       F5      &       6440    &       4.13    &$      0.10    \pm     0.09    $&$     3.48    ^{+     2.86    }_{     -1.36   }$&     Ic      &$      1.41    \pm     0.08    $&$     7.30    \pm     0.20    $&      WTTS    &$      6.61    \pm     0.06    $&      1,2,4   \\
RXJ1515.7-3332  &       K0      &       5250    &       1.68    &$      0.37    \pm     0.18    $&$     1.40    ^{+     1.35    }_{     -0.61   }$&     Ic      &$      1.22    \pm     0.30    $&$     7.22    \pm     0.36    $&      WTTS    &$      6.58    \pm     0.19    $&      1,2,3   \\
                                                                                                                                                                                                                                                                                        
RXJ1525.6-3537  &       K6      &       4205    &       0.14    &$      0.24    \pm     0.15    $&$     0.51    ^{+     0.49    }_{     -0.22   }$&     Ic      &$      0.95    \pm     0.11    $&$     6.88    \pm     0.54    $&      WTTS    &$      6.52    \pm     0.10    $&      1,2,4   \\
RXJ1527.3-3603  &       K7-M0   &       3955    &       0.12    &$      0.50    \pm     0.21    $&$     0.27    ^{+     0.28    }_{     -0.12   }$&     V       &$      0.70    \pm     0.11    $&$     6.98    \pm     0.60    $&      WTTS    &$      6.45    \pm     0.12    $&      1,2     \\
RXJ1529.3-3737  &       M3      &       3470    &       0.24    &0.00:&$        0.36    ^{+     0.24    }_{     -0.12   }$&     Ic      &$      0.35    \pm     0.06    $&$     6.24    \pm     0.19    $&      WTTS    &$      6.28    \pm     0.14    $&      1,2,4   \\
RXJ1529.7-3628  &       K2      &       4900    &       0.89    &$      0.46    \pm     0.20    $&$     1.89    ^{+     1.91    }_{     -0.84   }$&     V       &$      1.54    \pm     0.34    $&$     6.81    \pm     0.41    $&      WTTS    &$      6.63    \pm     0.17    $&      1,2,3   \\
Sz65*   &       M0      &       3850    &       19.40   &$      0.29    \pm     0.11    $&$     1.06    ^{+     1.07    }_{     -0.46   }$&     Ic      &$      0.56    \pm     0.08    $&$     6.00    \pm     0.36    $&      CTTS    &$      6.39    \pm     0.12    $&      2,6,10  \\
RXJ1539.7-3450  &       K4      &       4590    &       0.16    &$      0.50    \pm     0.21    $&$     0.54    ^{+     0.58    }_{     -0.25   }$&     V       &$      1.04    \pm     0.25    $&$     7.24    \pm     0.49    $&      WTTS    &$      6.54    \pm     0.18    $&      1,2     \\
                                                                                                                                                                                                                                                                                        
RXJ1540.3-3426A &       M3.5    &       3420    &       3.03    &0.00:&$        0.19    ^{+     0.16    }_{     -0.07   }$&     Ic      &$      0.31    \pm     0.05    $&$     6.48    \pm     0.27    $&      WTTS    &$      6.25    \pm     0.14    $&      1,2,11  \\
SSTc2dJ154148.3-350145  &       M2.7    &       3500    &       ---     &0.00:&$        0.23    ^{+     0.22    }_{     -0.09   }$&     Ic      &$      0.36    \pm     0.07    $&$     6.45    \pm     0.30    $&      WTTS    &$      6.28    \pm     0.15    $&      2,11,12 \\
RXJ1542.0-3601  &       K7      &       4060    &       0.32    &$      0.23    \pm     0.18    $&$     0.20    ^{+     0.20    }_{     -0.09   }$&     V       &$      0.76    \pm     0.12    $&$     7.34    \pm     0.44    $&      WTTS    &$      6.46    \pm     0.12    $&      1,2     \\
RXJ1544.0-3311* &       K0      &       5250    &       0.87    &$      0.65    \pm     0.19    $&$     1.78    ^{+     1.73    }_{     -0.78   }$&     V       &$      1.33    \pm     0.36    $&$     7.11    \pm     0.37    $&      WTTS    &$      6.60    \pm     0.20    $&      1,2,3   \\
RXJ1546.6-3618  &       K1      &       5080    &       0.67    &$      0.58    \pm     0.33    $&$     1.62    ^{+     2.10    }_{     -0.83   }$&     Ic      &$      1.38    \pm     0.46    $&$     7.03    \pm     0.45    $&      WTTS    &$      6.61    \pm     0.25    $&      1,2,3   \\
RXJ1546.7-3459  &       M0      &       3850    &       1.53    &0.00:&$        0.29    ^{+     0.24    }_{     -0.11   }$&     J       &$      0.59    \pm     0.09    $&$     6.74    \pm     0.49    $&      WTTS    &$      6.40    \pm     0.12    $&      1,2     \\
RXJ1547.1-3540  &       K5      &       4350    &       0.55    &$      0.20    \pm     0.18    $&$     0.24    ^{+     0.24    }_{     -0.10   }$&     V       &$      0.80    \pm     0.10    $&$     7.61    \pm     0.31    $&      WTTS    &$      6.48    \pm     0.10    $&      1,2     \\
RXJ1547.6-4018  &       K1      &       5080    &       1.60    &$      0.35    \pm     0.07    $&$     1.08    ^{+     0.83    }_{     -0.40   }$&     Ic      &$      1.17    \pm     0.28    $&$     7.24    \pm     0.33    $&      WTTS    &$      6.57    \pm     0.18    $&      1,2,3   \\
HMLup   &       M3      &       3470    &       115.30  &0.00:&$        0.26    ^{+     0.25    }_{     -0.10   }$&     Ic      &$      0.34    \pm     0.08    $&$     6.37    \pm     0.27    $&      CTTS    &$      6.27    \pm     0.17    $&      2,6     \\
HNLup*  &       M1.5    &       3650    &       49.60   &$      2.14    \pm     0.11    $&$     1.32    ^{+     2.67    }_{     -0.71   }$&     Ic      &$      0.42    \pm     0.09    $&$     5.88    \pm     0.25    $&      CTTS    &$      6.32    \pm     0.17    $&      2,6,10  \\
RXJ1548.1-3452  &       M2.5    &       3525    &       2.75    &0.00:&$        0.34    ^{+     0.24    }_{     -0.12   }$&     Ic      &$      0.37    \pm     0.05    $&$     6.29    \pm     0.24    $&      WTTS    &$      6.29    \pm     0.12    $&      1,2,4   \\
                                                                                                                                                                                                                                                                                        
RXJ1548.9-3513  &       K6      &       4205    &       0.16    &$      0.51    \pm     0.18    $&$     0.15    ^{+     0.15    }_{     -0.07   }$&     V       &$      0.70    \pm     0.11    $&$     7.79    \pm     0.44    $&      WTTS    &$      6.44    \pm     0.12    $&      1,2     \\
Sz76    &       M1      &       3720    &       10.30   &$      0.38    \pm     0.18    $&$     0.19    ^{+     0.20    }_{     -0.09   }$&     Ic      &$      0.47    \pm     0.05    $&$     6.80    \pm     0.55    $&      WTTS    &$      6.35    \pm     0.10    $&      2,6,8   \\
HD141277*       &       K1      &       5080    &       1.52    &$      0.49    \pm     0.34    $&$     2.25    ^{+     2.99    }_{     -1.17   }$&     Ic      &$      1.57    \pm     0.54    $&$     6.87    \pm     0.46    $&      WTTS    &$      6.64    \pm     0.25    $&      1,2,3   \\
RXJ1550.7-3828  &       K7      &       4060    &       0.88    &$      0.86    \pm     0.30    $&$     1.10    ^{+     1.37    }_{     -0.55   }$&     Ic      &$      0.75    \pm     0.12    $&$     6.20    \pm     0.47    $&      WTTS    &$      6.46    \pm     0.13    $&      1,2,4   \\
Sz77*   &       M0      &       3850    &       12.40   &0.00:&$        0.62    ^{+     0.46    }_{     -0.22   }$&     Ic      &$      0.57    \pm     0.08    $&$     6.27    \pm     0.38    $&      CTTS    &$      6.39    \pm     0.12    $&      2,6,10  \\
RXJ1555.4-3338  &       K5      &       4350    &       ---     &0.00:&$        0.34    ^{+     0.23    }_{     -0.12   }$&     Ic      &$      0.91    \pm     0.15    $&$     7.35    \pm     0.41    $&      WTTS    &$      6.51    \pm     0.13    $&      1,2,4   \\
                                                                                                                                                                                                                                                                                        
RXJ1556.0-3655  &       M1      &       3720    &       73.60   &$      0.14    \pm     0.07    $&$     0.47    ^{+     0.42    }_{     -0.19   }$&     Ic      &$      0.48    \pm     0.05    $&$     6.28    \pm     0.37    $&      CTTS    &$      6.35    \pm     0.10    $&      1,2,4   \\
Sz82    *&      M0      &       3850    &       39.00   &0.00:&$        0.93    ^{+     0.84    }_{     -0.36   }$&     Ic      &$      0.56    \pm     0.08    $&$     6.06    \pm     0.35    $&      CTTS    &$      6.39    \pm     0.12    $&      2,6,8,10        \\
                                                                                                                                                                                                                                                                                        
Sz126*  &       K7-M0   &       3955    &       ---     &$      0.60    \pm     0.13    $&$     0.43    ^{+     0.41    }_{     -0.18   }$&     Ic      &$      0.68    \pm     0.12    $&$     6.66    \pm     0.54    $&      CTTS    &$      6.44    \pm     0.14    $&      2,6,10,13       \\
                                                                                                                                                                                                                                                                                        
Sz128*  &       M1.5    &       3650    &       9.90    &$      1.86    \pm     0.13    $&$     0.36    ^{+     0.41    }_{     -0.17   }$&     Ic      &$      0.43    \pm     0.10    $&$     6.36    \pm     0.41    $&      CTTS    &$      6.33    \pm     0.17    $&      2,6,14  \\
RXJ1558.9-3646  &       M1.5    &       3650    &       2.81    &0.00:&$        0.46    ^{+     0.36    }_{     -0.17   }$&     Ic      &$      0.43    \pm     0.10    $&$     6.24    \pm     0.31    $&      WTTS    &$      6.33    \pm     0.17    $&      1,2,4   \\
CD-3610569      &       K3      &       4730    &       0.71    &$      0.58    \pm     0.08    $&$     0.69    ^{+     0.54    }_{     -0.26   }$&     Ic      &$      1.10    \pm     0.27    $&$     7.24    \pm     0.39    $&      WTTS    &$      6.55    \pm     0.18    $&      1,2,3   \\
RXJ1559.9-3750  &       M0.5    &       3785    &       0.51    &$      0.32    \pm     0.28    $&$     1.01    ^{+     1.19    }_{     -0.49   }$&     Ic      &$      0.51    \pm     0.10    $&$     5.98    \pm     0.47    $&      WTTS    &$      6.37    \pm     0.15    $&      1,2,4   \\
SSTc2dJ160000.6-422158  &       M3.7    &       3400    &       ---     &0.00:&$        0.15    ^{+     0.12    }_{     -0.05   }$&     Ic      &$      0.30    \pm     0.05    $&$     6.57    \pm     0.31    $&      WTTS    &$      6.24    \pm     0.13    $&      2,11,12,15      \\
Sz131   &       M2      &       3580    &       31.90   &$      2.48    \pm     0.24    $&$     0.55    ^{+     0.96    }_{     -0.30   }$&     Ic      &$      0.41    \pm     0.06    $&$     6.12    \pm     0.77    $&      CTTS    &$      6.31    \pm     0.12    $&      2,6,15  \\
RXJ1601.9-3613  &       K3      &       4730    &       0.45    &$      0.59    \pm     0.21    $&$     0.88    ^{+     0.97    }_{     -0.41   }$&     Ic      &$      1.20    \pm     0.29    $&$     7.10    \pm     0.48    $&      WTTS    &$      6.57    \pm     0.18    $&      1,2,4   \\
EXLup   &       M0      &       3850    &       43.30   &0.00:&$        1.09    ^{+     0.85    }_{     -0.40   }$&     Ic      &$      0.56    \pm     0.08    $&$     5.99    \pm     0.31    $&      CTTS    &$      6.39    \pm     0.12    $&      2,6     \\
RXJ1603.8-3938* &       K3      &       4730    &       1.13    &$      0.43    \pm     0.08    $&$     0.55    ^{+     0.44    }_{     -0.21   }$&     Ic      &$      1.01    \pm     0.25    $&$     7.35    \pm     0.39    $&      WTTS    &$      6.53    \pm     0.19    $&      1,2,3,16        \\
HD143978        &       G2      &       5860    &       3.60    &0.00:&$        1.29    ^{+     0.94    }_{     -0.46   }$&     Ic      &$      1.11    \pm     0.07    $&$     7.88    \pm     0.48    $&      WTTS    &$      6.56    \pm     0.07    $&      1,2,4   \\
RXJ1605.5-3837  &       M1      &       3720    &       1.28    &0.00:&$        0.21    ^{+     0.16    }_{     -0.08   }$&     Ic      &$      0.50    \pm     0.05    $&$     6.74    \pm     0.42    $&      WTTS    &$      6.36    \pm     0.09    $&      1,2,4   \\
HOLup*  &       M1      &       3720    &       219.80  &0.00:&$        0.57    ^{+     0.46    }_{     -0.21   }$&     Ic      &$      0.48    \pm     0.05    $&$     6.19    \pm     0.31    $&      CTTS    &$      6.35    \pm     0.10    $&      2,6,10  \\
Sz90*   &       K7-M0   &       3955    &       28.50   &$      2.12    \pm     0.40    $&$     0.80    ^{+     1.48    }_{     -0.47   }$&     Ic      &$      0.66    \pm     0.13    $&$     6.26    \pm     0.65    $&      CTTS    &$      6.43    \pm     0.15    $&      2,6,10  \\
Sz91*   &       M0.5    &       3785    &       95.90   &$      1.93    \pm     0.11    $&$     0.36    ^{+     0.35    }_{     -0.15   }$&     Ic      &$      0.52    \pm     0.10    $&$     6.51    \pm     0.44    $&      CTTS/TD &$      6.38    \pm     0.15    $&      2,6,10  \\
RXJ1607.2-3839  &       K7      &       4060    &       1.40    &$      0.29    \pm     0.20    $&$     0.93    ^{+     1.04    }_{     -0.43   }$&     Ic      &$      0.75    \pm     0.11    $&$     6.29    \pm     0.47    $&      WTTS    &$      6.46    \pm     0.12    $&      1,2,4   \\
Sz95    *&      M1.5    &       3650    &       10.20   &$      1.29    \pm     0.11    $&$     0.40    ^{+     0.39    }_{     -0.17   }$&     Ic      &$      0.43    \pm     0.10    $&$     6.30    \pm     0.37    $&      CTTS/TD &$      6.33    \pm     0.17    $&      2,6,15  \\
RXJ1608.0-3857  &       M0      &       3850    &       1.14    &$      0.24    \pm     0.22    $&$     0.92    ^{+     1.03    }_{     -0.43   }$&     Ic      &$      0.56    \pm     0.08    $&$     6.06    \pm     0.42    $&      WTTS    &$      6.39    \pm     0.12    $&      1,2,4   \\
Sz96*   &       M1.5    &       3650    &       11.00   &$      0.43    \pm     0.11    $&$     0.53    ^{+     0.50    }_{     -0.22   }$&     Ic      &$      0.44    \pm     0.10    $&$     6.17    \pm     0.33    $&      CTTS/TD &$      6.33    \pm     0.17    $&      2,6,10  \\
RXJ1608.3-3843  &       K7      &       4060    &       0.11    &0.00:&$        0.39    ^{+     0.28    }_{     -0.14   }$&     Ic      &$      0.80    \pm     0.12    $&$     6.87    \pm     0.44    $&      WTTS    &$      6.48    \pm     0.12    $&      1,2,4   \\
Sz97    &       M3      &       3470    &       58.20   &$      0.95    \pm     0.11    $&$     0.28    ^{+     0.27    }_{     -0.12   }$&     Ic      &$      0.34    \pm     0.08    $&$     6.34    \pm     0.28    $&      CTTS    &$      6.28    \pm     0.17    $&      2,6     \\

RXJ1608.6-3922  &       K6      &       4205    &       7.42    &$      1.52    \pm     0.07    $&$     0.88    ^{+     0.74    }_{     -0.34   }$&     Ic      &$      0.93    \pm     0.13    $&$     6.49    \pm     0.47    $&      CTTS    &$      6.51    \pm     0.11    $&      1,2,4,8 \\

SSTc2dJ160853.2-391440  &       K7.8    &       3900    &       ---     &$      4.60    \pm     0.43    $&$     0.51    ^{+     0.95    }_{     -0.30   }$&     Ic      &$      0.61    \pm     0.10    $&$     6.45    \pm     0.70    $&      CTTS    &$      6.41    \pm     0.13    $&      2,11,12,15      \\
RXJ1608.9-3905  &       K2      &       4900    &       0.84    &$      0.38    \pm     0.18    $&$     3.32    ^{+     3.71    }_{     -1.54   }$&     Ic      &$      1.86    \pm     0.34    $&$     6.47    \pm     0.40    $&      WTTS    &$      6.68    \pm     0.14    $&      1,2,4   \\
RXJ1608.9-3945  &       M3.5    &       3420    &       2.69    &$      0.16    \pm     0.14    $&$     0.25    ^{+     0.25    }_{     -0.11   }$&     Ic      &$      0.32    \pm     0.06    $&$     6.36    \pm     0.28    $&      WTTS    &$      6.26    \pm     0.15    $&      1,2,4   \\
Sz111   &       M1.5    &       3650    &       145.20  &0.00:&$        0.34    ^{+     0.27    }_{     -0.13   }$&     Ic      &$      0.43    \pm     0.10    $&$     6.38    \pm     0.33    $&      CTTS    &$      6.33    \pm     0.17    $&      2,6     \\
Sz112   &       M4      &       3370    &       46.70   &$      0.81    \pm     0.11    $&$     0.30    ^{+     0.31    }_{     -0.13   }$&     Ic      &$      0.30    \pm     0.05    $&$     6.26    \pm     0.29    $&      CTTS/TD &$      6.24    \pm     0.14    $&      2,6     \\
                                                                                                                                                                                                                                                                                        
V908Sco &       M4      &       3370    &       33.00   &0.00:&$        0.48    ^{+     0.37    }_{     -0.17   }$&     Ic      &$      0.30    \pm     0.04    $&$     6.18    \pm     0.55    $&      CTTS    &$      6.24    \pm     0.13    $&      2,6     \\
                                                                                                                                                                                                                                                                                        
Sz115   &       M4      &       3370    &       7.10    &0.00:&$        0.15    ^{+     0.13    }_{     -0.06   }$&     Ic      &$      0.29    \pm     0.05    $&$     6.55    \pm     0.30    $&      WTTS    &$      6.23    \pm     0.15    $&      2,6,15,19       \\
Sz134   &       M1      &       3720    &       90.50   &0.00:&$        0.25    ^{+     0.18    }_{     -0.09   }$&     Ic      &$      0.47    \pm     0.05    $&$     6.64    \pm     0.40    $&      CTTS    &$      6.35    \pm     0.10    $&      2,6     \\
RXJ1609.4-3850  &       M0.5    &       3785    &       1.64    &$      0.13    \pm     0.12    $&$     1.41    ^{+     1.34    }_{     -0.60   }$&     Ic      &$      0.51    \pm     0.08    $&$     5.86    \pm     0.32    $&      WTTS    &$      6.37    \pm     0.13    $&      1,2,4   \\
Sz116*  &       M1.5    &       3650    &       3.90    &$      0.42    \pm     0.24    $&$     0.30    ^{+     0.35    }_{     -0.15   }$&     Ic      &$      0.43    \pm     0.10    $&$     6.45    \pm     0.43    $&      WTTS    &$      6.33    \pm     0.18    $&      2,6,15  \\
Sz117   &       M2      &       3580    &       20.50   &$      0.43    \pm     0.11    $&$     0.30    ^{+     0.29    }_{     -0.13   }$&     Ic      &$      0.39    \pm     0.05    $&$     6.38    \pm     0.32    $&      CTTS    &$      6.31    \pm     0.12    $&      2,6     \\
Sz118   &       K6      &       4205    &       28.30   &$      4.19    \pm     0.11    $&$     0.62    ^{+     2.57    }_{     -0.38   }$&     Ic      &$      0.96    \pm     0.14    $&$     6.73    \pm     0.88    $&      CTTS    &$      6.52    \pm     0.11    $&      2,6     \\
RXJ1609.9-3923* &       M1.5    &       3650    &       13.67   &$      2.33    \pm     0.07    $&$     1.33    ^{+     1.24    }_{     -0.54   }$&     Ic      &$      0.42    \pm     0.09    $&$     5.88    \pm     0.28    $&      CTTS/TD &$      6.32    \pm     0.17    $&      1,2,4   \\
Sz119   &       M4      &       3370    &       4.50    &$      0.53    \pm     0.18    $&$     0.39    ^{+     0.41    }_{     -0.18   }$&     Ic      &$      0.30    \pm     0.05    $&$     6.18    \pm     0.61    $&      WTTS    &$      6.25    \pm     0.13    $&      2,6     \\
                                                                                                                                                                                                                                                                                        
Sz122   &       M2      &       3580    &       4.80    &$      0.37    \pm     0.18    $&$     0.26    ^{+     0.28    }_{     -0.12   }$&     Ic      &$      0.39    \pm     0.05    $&$     6.45    \pm     0.38    $&      WTTS    &$      6.31    \pm     0.12    $&      2,6     \\
Sz123*  &       M1      &       3720    &       250.60  &$      0.86    \pm     0.11    $&$     0.39    ^{+     0.41    }_{     -0.17   }$&     Ic      &$      0.48    \pm     0.05    $&$     6.38    \pm     0.43    $&      CTTS    &$      6.35    \pm     0.10    $&      2,6,10  \\
RXJ1612.0-3840  &       K5      &       4350    &       0.72    &$      0.61    \pm     0.26    $&$     3.62    ^{+     4.14    }_{     -1.74   }$&     Ic      &$      1.15    \pm     0.25    $&$     5.87    \pm     0.41    $&      WTTS    &$      6.56    \pm     0.16    $&      1,2,4   \\
SSTc2dJ161207.6-381324  &       M4.5    &       3305    &       3.60    &0.00:&$        0.51    ^{+     0.37    }_{     -0.18   }$&     Ic      &$      0.25    \pm     0.05    $&$     6.14    \pm     0.64    $&      WTTS    &$      6.20    \pm     0.15    $&      2,11,12,18      \\
                                                                                                                                                                                                                                                                                        
SSTc2dJ161243.8-381503  &       K4.8    &       4400    &       ---     &$      1.27    \pm     0.12    $&$     2.19    ^{+     2.08    }_{     -0.92   }$&     Ic      &$      1.20    \pm     0.18    $&$     6.18    \pm     0.45    $&      CTTS    &$      6.57    \pm     0.12    $&      2,11,12,15      \\
RXJ1614.4-3808  &       K7      &       4060    &       0.46    &$      0.23    \pm     0.17    $&$     0.15    ^{+     0.14    }_{     -0.06   }$&     V       &$      0.70    \pm     0.11    $&$     7.54    \pm     0.35    $&      WTTS    &$      6.45    \pm     0.12    $&      1,2     \\
HD147402        &       G3      &       5830    &       ---     &$      0.37    \pm     0.28    $&$     1.68    ^{+     1.93    }_{     -0.81   }$&     Ic      &$      1.14    \pm     0.16    $&$     7.40    \pm     0.16    $&      WTTS    &$      6.56    \pm     0.11    $&      2,3,8   \\

\end{longtable}
}
\tablefoot{
For each star, we provide the spectral type, temperature, H$\alpha$ equivalent
width, visual extinction, luminosity, photometric filter used to compute luminosities, the mass and age derived from the \citet{Siess(2000)} models, disk lifetime computed from Eq.~(5), T~Tauri subclass, and sources of photometric/spectroscopic information. The $A_{V}$ estimate of 0.00: denotes a non-physical negative value or a non-significant result. Stars marked with ``*'' are binary (or multiple) systems (see Sect.~2).
}

\tablebib{
(1)~\citet{Krautter(1997)};
(2)~2MASS \citep{Cutri(2003)};
(3)~\citet{Torres(2006)};
(4)~\citet{Wichmann(1997a)};
(5)~\citet{Sartori(2003)};
(6)~\citet{Hughes(1994)};
(7)~\citet{Neuhauser(2008)};
(8)~\citet{Wahhaj(2010)};
(9)~\citet{Wichmann(1997b)};
(10)~\citet{Ghez(1997)};
(11)~\citet{Comeron(2009)};
(12)~\citet{Marti(2011)};
(13)~\citet{Valenti(2003)};
(14)~\citet{Brandner(1996)};
(15)~\citet{Merin(2008)};
(16)~\citet{Guenther(2001)};
(17)~\citet{Alcala(2014)};
(18)~\citet{Comeron(2013)};
(19)~\citet{Cieza(2007)};
}

\end{longtab}

%%%%%%%%%%%%%%%%%%%%%%%%%%%%%%%%%%%%%%%%%%% 
%                                                       FIGURE 2
%%%%%%%%%%%%%%%%%%%%%%%%%%%%%%%%%%%%%%%%%%% 
\begin{figure*}[!btp]
\begin{center}
\includegraphics[width=0.95\textwidth]{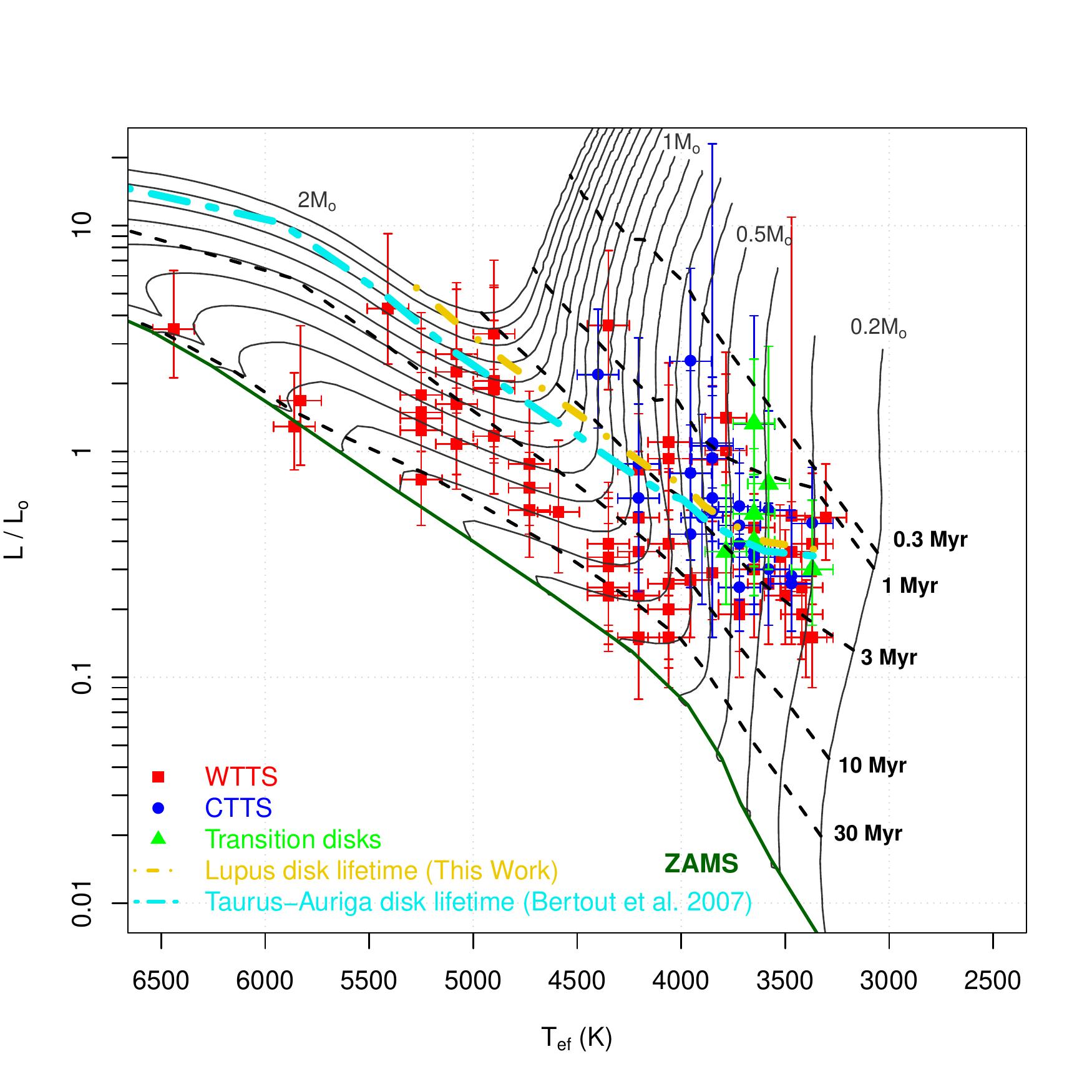}
\caption{
\label{HRD}
HRD of the Lupus association. Red squares and blue circles denote, respectively, the WTTSs and CTTSs in our sample. The green triangles represent those stars with transition disks (see Sect.~4). The error bars indicate the $1\sigma$ uncertainties on the stellar luminosities and effective temperatures. The black solid lines and dashed lines indicate, respectively, the evolutionary tracks and isochrones computed by \citet{Siess(2000)} with $Y=0.277$ and $Z=0.02$. The green solid line denotes the zero-age main sequence (ZAMS) computed with that model. The dash-dotted lines show the disk lifetime $\tau_{d}$ for the TTSs in the Lupus (this work) and Taurus-Auriga \citep{Bertout(2007)} associations. }
\end{center}
\end{figure*}

%%%%%%%%%%%%%%%%%%%%%%%%%%%%%%%%%%%%%%%%%%% 
%                                                       SECTION 3
%%%%%%%%%%%%%%%%%%%%%%%%%%%%%%%%%%%%%%%%%%% 
\section{Revisiting the CTTSs and WTTSs in Lupus}

In the following we use the stellar luminosities and effective temperatures of the stars to place them in the HRD and discuss the age and mass distributions of the Lupus association. We derive the mass and age of the stars by interpolating between the evolutionary tracks and isochrones of the pre-main sequence stars models computed by \citet{Siess(2000)}. Among the various models that exist in the literature, we chose the \citet{Siess(2000)}  grid of pre-main sequence stars  to allow for a direct comparison with the results in the Taurus-Auriga association (see Sect.~5) obtained by \citet{Bertout(2007)}. The uncertainties affecting our mass and age estimates are estimated by defining the upper and lower limits for the stellar luminosities and effective temperatures within their error bars, and interpolating between the tracks and isochrones. We assumed an error of $\pm100$~K on the effective temperatures that roughly corresponds to an error of one subtype of spectral type for the late-type stars in our sample. Our mass and age results are presented in Table~1.

The HRD of the Lupus association illustrated in Fig.~\ref{HRD} shows that a few WTTSs and CTTSs share the same region of the HRD. As already discussed in Paper~I (see, e.g., Fig.~15 of that paper), the WTTSs in our sample comprise an \textit{\emph{on-cloud}} population, which is located in the direction of the main star-forming clouds (Lupus 1-4) of the Lupus SFR, and a more dispersed \textit{\emph{off-cloud}} population surrounding the clouds. The question arises whether the two populations of WTTSs and the CTTSs of the Lupus association exhibit the same mass and age distributions. Table~2 lists the average mass and age of each population for comparison and reveals the different properties of the CTTSs and WTTSs in our sample. 

%%%%%%%%%%%%%%%%%%%%%%%%%%%%%%%%%%%%%%%%%%% 
%                                                       TABLE 2
%%%%%%%%%%%%%%%%%%%%%%%%%%%%%%%%%%%%%%%%%%% 
\begin{table}[!h]
\renewcommand\thetable{2} 
\centering
\caption{
\label{tab2}
Mean masses and ages of the TTSs in the Lupus association derived from the \citet{Siess(2000)} models. 
}
\resizebox{9cm}{!} {
\begin{tabular}{lccccc}

%\hline
%\hline
%&\multicolumn{4}{c}{\textbf{\citet{Siess(2000)}}}\\
%\hline
\hline
Sample&Stars&Mean&Median&Mean&Median\\
&&$M/M_{\odot}$&$M/M_{\odot}$&$\log t$&$\log t$\\
&&&&(t in Myr)&(t in Myr)\\
\hline
CTTS&30&$0.54\pm0.04$&0.48&$6.32\pm0.24$&6.28\\
WTTS \footnotesize{(on-cloud)}&33&$0.69\pm0.07$&0.59&$7.05\pm0.67$&6.55\\
WTTS \footnotesize{(off-cloud)}&29&$1.09\pm0.07$&1.10&$7.19\pm0.28$&7.11\\
\hline
WTTS \footnotesize{(all stars)}&62&$0.88\pm0.05$&0.81&$7.12\pm0.48$&6.87\\
\hline

\end{tabular}
}
\end{table}

We computed our mass and age estimates again by using the \citet{DM97} tracks and the \citet{Baraffe(1998)} models to demonstrate that the different age and mass distributions are not an artifact from our choice of evolutionary model (see Table~3). Although evolutionary models computed by different groups yield different mass and age estimates for a given star, we note from Tables~2 and 3 that the results obtained with the various models employed in this work are consistent between themselves.  This allows us to confirm that (i) the CTTSs are on average younger than the WTTSs, and (ii) the \textit{\emph{off-cloud}} stars in the Lupus association are on average older than the \textit{\emph{on-cloud}} stars.

%%%%%%%%%%%%%%%%%%%%%%%%%%%%%%%%%%%%%%%%%%% 
%                                                       TABLE 3
%%%%%%%%%%%%%%%%%%%%%%%%%%%%%%%%%%%%%%%%%%% 
\begin{table*}[!btp]
\renewcommand\thetable{3} 
\centering
\caption{
\label{tab3}
Mean masses and ages of the TTSs in the Lupus association derived from the \citet{DM97} and \citet{Baraffe(1998)} pre-main sequence stars evolutionary models. 
}
\resizebox{17cm}{!} {
\begin{tabular}{lcccccccc}

\hline
\hline
&\multicolumn{4}{c}{\textbf{\citet{DM97}}}&\multicolumn{4}{c}{\textbf{\citet{Baraffe(1998)}}}\\
\hline
\hline
Sample&Mean&Median&Mean&Median&Mean&Median&Mean&Median\\
&$M/M_{\odot}$&$M/M_{\odot}$&$\log t$&$\log t$&$M/M_{\odot}$&$M/M_{\odot}$&$\log t$&$\log t$\\
&&&(t in Myr)&(t in Myr)&&&(t in Myr)&(t in Myr)\\
\hline
CTTS&$0.40\pm0.02$&0.38&$6.15\pm0.21$&6.10&$0.63\pm0.04$&0.60&$6.38\pm0.20$&6.35\\
WTTS \footnotesize{(on-cloud)}&$0.61\pm0.06$&0.47&$6.79\pm0.59$&6.42&$0.72\pm0.06$&0.73&$7.05\pm0.45$&6.80\\
WTTS \footnotesize{(off-cloud)}&$1.00\pm0.07$&1.00&$6.93\pm0.35$&6.79&$0.99\pm0.05$&0.98&$7.17\pm0.28$&7.09\\
\hline
WTTS \footnotesize{(all stars)}&$0.80\pm0.05$&0.80&$6.86\pm0.46$&6.63&$0.85\pm0.04$&0.86&$7.11\pm0.36$&7.04\\
\hline

\end{tabular}
}
\end{table*}

We performed a Kolmogorov-Smirnov (KS) and a Wilcoxon rank sum test to investigate the null hypothesis that the age and mass distributions of the CTTSs and WTTSs in the Lupus association derive from the same distribution. The results of this analysis are given in Tables~4 and 5. Adopting the significance level of $\alpha=0.05$ \citep[see, e.g.,][]{Feigelson(2012)}, we conclude that the null hypothesis can be rejected in each case regarding stellar ages (see Table~4), which means that the age distributions of the CTTSs and WTTSs in Lupus are not identical. On the other hand, our results with the statistical tests applied to the mass distribution of CTTSs and  \textit{\emph{on-cloud}} WTTSs are marginally above the adopted significance level in a few cases (see Table~5) and remain inconclusive.

%%%%%%%%%%%%%%%%%%%%%%%%%%%%%%%%%%%%%%%%%%% 
%                                                       TABLE 4 
%%%%%%%%%%%%%%%%%%%%%%%%%%%%%%%%%%%%%%%%%%% 

\begin{table*}[!htp]
\renewcommand\thetable{4} 
\centering
\caption{
\label{tab4}
Results of the Kolmogorov-Smirnov (KS) and Wilcoxon rank sum (WRS) statistical tests applied to the age distributions of the TTSs in the Lupus association computed with different sets of isochrones. We provide the \textit{\emph{p-value}} given by each test. 
}
\resizebox{18cm}{!} {
\begin{tabular}{lccccccccc}

\hline
\hline
&&\multicolumn{2}{c}{\textbf{\citet{Siess(2000)}}}&\multicolumn{2}{c}{\textbf{\citet{DM97}}}&\multicolumn{2}{c}{\textbf{\citet{Baraffe(1998)}}}\\
\hline
\hline
Sample 1&Sample 2&$p_{KS}$&$p_{WRS}$&$p_{KS}$&$p_{WRS}$&$p_{KS}$&$p_{WRS}$\\
\hline
CTTS&WTTS \footnotesize{(on-cloud)}&$2.13\times 10^{-3}$&$1.31\times 10^{-3}$&$3.06\times 10^{-4}$&$3.52\times 10^{-4}$&$1.65\times 10^{-3}$&$2.71\times 10^{-4}$\\
CTTS&WTTS \footnotesize{(off-cloud)}&$1.01\times 10^{-10}$&$2.01\times 10^{-9}$&$5.66\times 10^{-9}$&$4.27\times 10^{-8}$&$3.27\times 10^{-7}$&$1.24\times 10^{-5}$\\
CTTS&WTTS \footnotesize{(all stars)}&$4.19\times 10^{-8}$&$1.57\times 10^{-7}$&$3.95\times 10^{-8}$&$2.31\times 10^{-7}$&$1.66\times 10^{-6}$&$4.25\times 10^{-6}$\\
\hline

\end{tabular}
}
\end{table*}

%%%%%%%%%%%%%%%%%%%%%%%%%%%%%%%%%%%%%%%%%%% 
%                                                       TABLE 5 
%%%%%%%%%%%%%%%%%%%%%%%%%%%%%%%%%%%%%%%%%%% 

\begin{table*}[!htp]
\renewcommand\thetable{5} 
\centering
\caption{
\label{tab5}
Results of the Kolmogorov-Smirnov (KS) and Wilcoxon rank sum (WRS) statistical tests applied to the mass distributions of the TTSs in the Lupus association computed with different evolutionary tracks. We provide the \textit{\emph{p-value}} given by each test. 
}
\resizebox{18cm}{!} {
\begin{tabular}{lccccccccc}

\hline
\hline
&&\multicolumn{2}{c}{\textbf{\citet{Siess(2000)}}}&\multicolumn{2}{c}{\textbf{\citet{DM97}}}&\multicolumn{2}{c}{\textbf{\citet{Baraffe(1998)}}}\\
\hline
\hline
Sample 1&Sample 2&$p_{KS}$&$p_{WRS}$&$p_{KS}$&$p_{WRS}$&$p_{KS}$&$p_{WRS}$\\
\hline
CTTS&WTTS \footnotesize{(on-cloud)}&$3.84\times 10^{-2}$&$2.13\times 10^{-1}$&$1.95\times 10^{-2}$&$7.72\times 10^{-2}$&$1.82\times 10^{-1}$&$2.95\times 10^{-1}$\\
CTTS&WTTS \footnotesize{(off-cloud)}&$1.42\times 10^{-8}$&$1.10\times 10^{-7}$&$9.25\times 10^{-10}$&$4.79\times 10^{-8}$&$9.88\times 10^{-6}$&$1.62\times 10^{-5}$\\
CTTS&WTTS \footnotesize{(all stars)}&$6.16\times 10^{-6}$&$2.39\times 10^{-4}$&$1.97\times 10^{-6}$&$4.20\times 10^{-5}$&$5.45\times 10^{-4}$&$2.28\times 10^{-3}$\\
\hline

\end{tabular}
}
\end{table*}

\begin{figure*}[!htp]
\begin{center}
\includegraphics[width=0.85\textwidth]{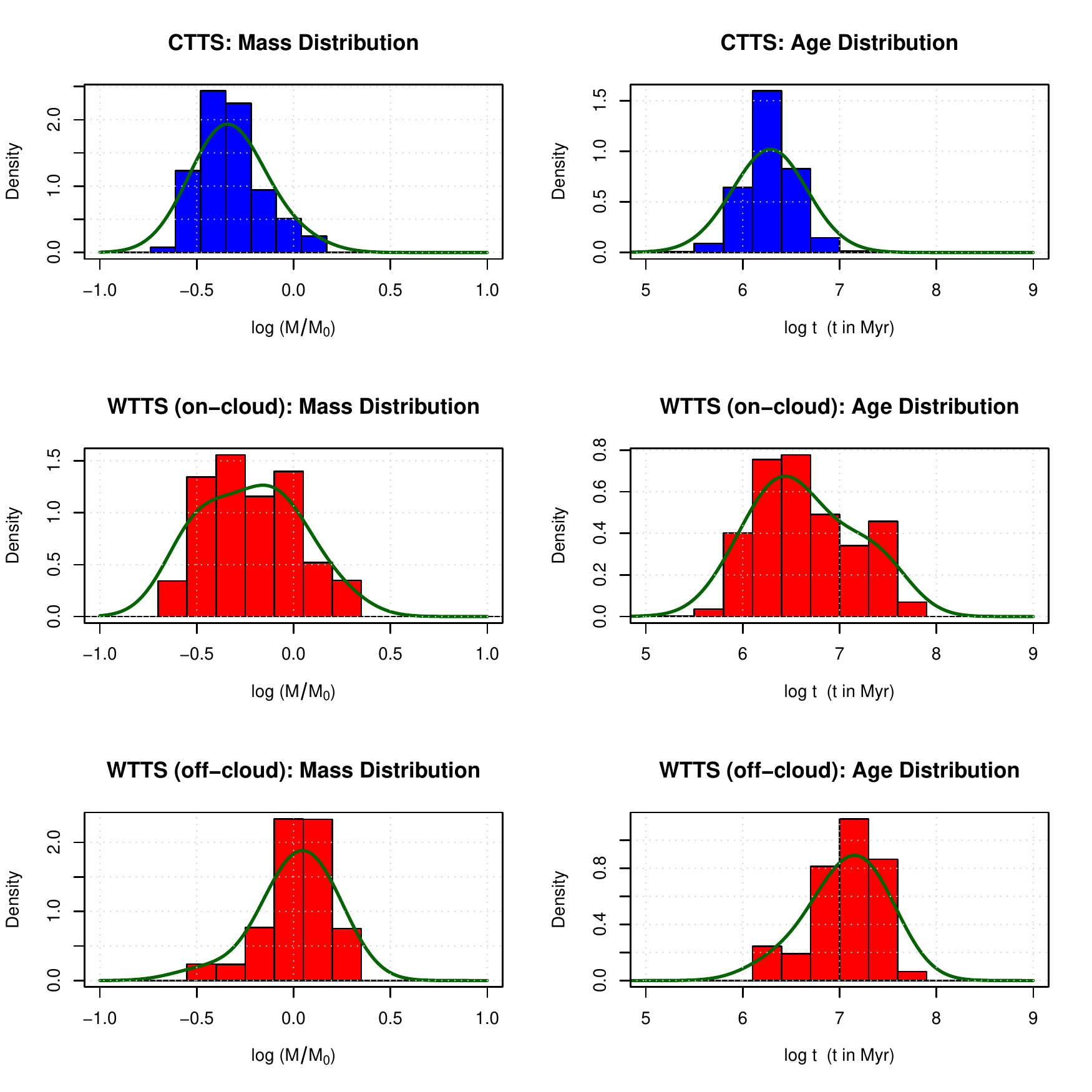}
\caption{
\label{age_mass_hist}
Mass and age distributions of the CTTSs and WTTSs of the Lupus association obtained after 1000 Monte Carlo simulations. The green solid line indicates the kernel density estimator.
 }
\end{center}
\end{figure*}

\begin{figure*}[!htp]
\begin{center}
\includegraphics[width=0.4\textwidth]{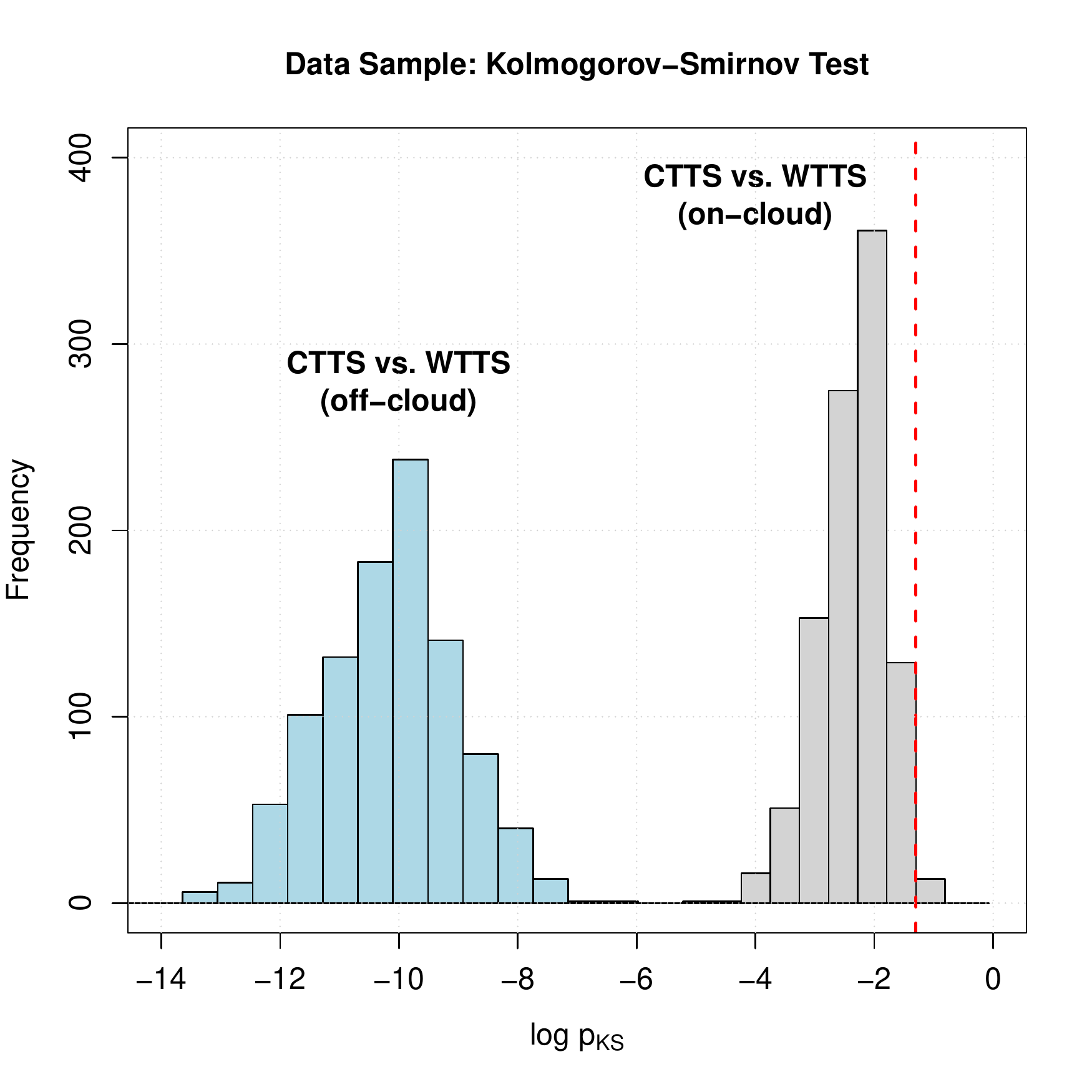}
\includegraphics[width=0.4\textwidth]{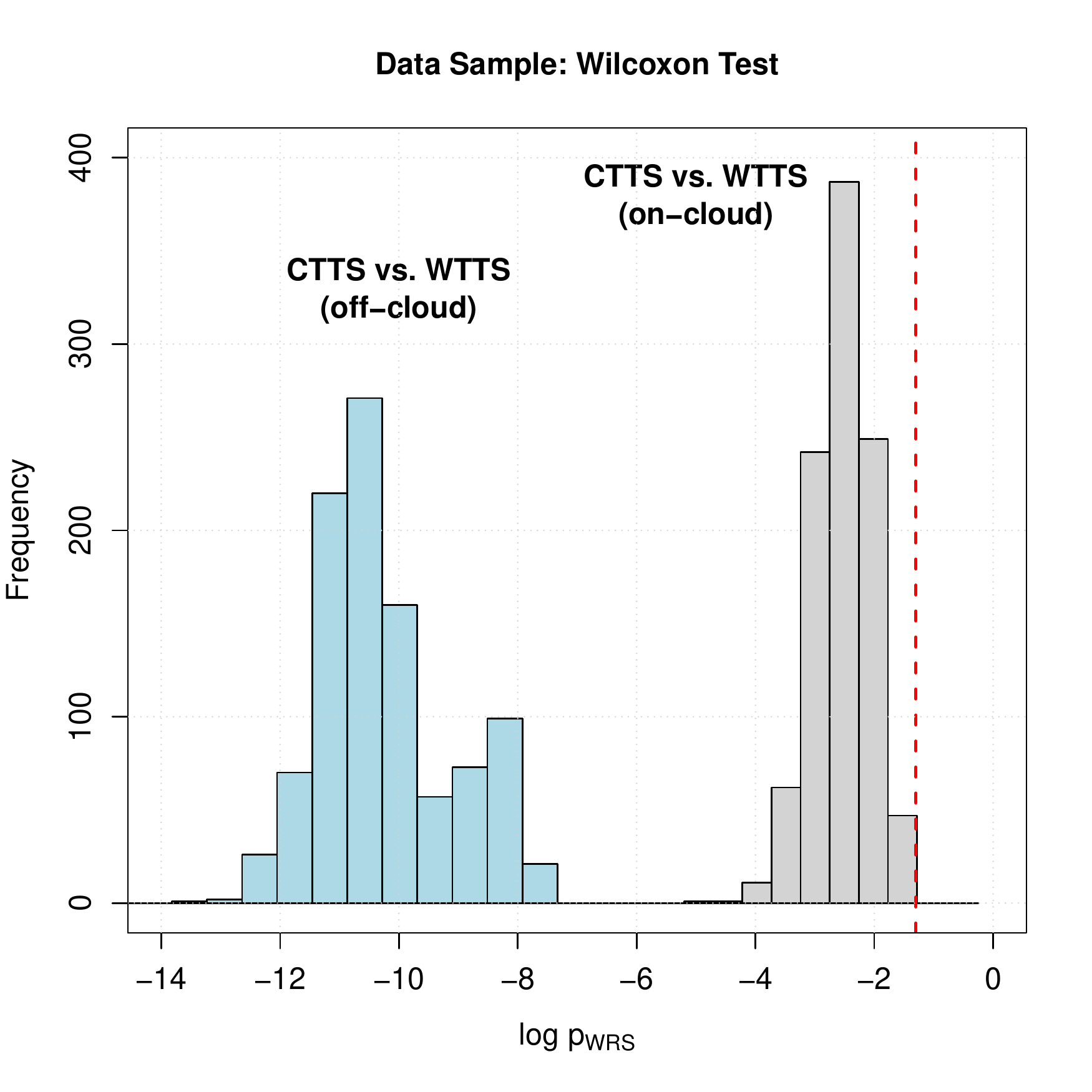}
\includegraphics[width=0.4\textwidth]{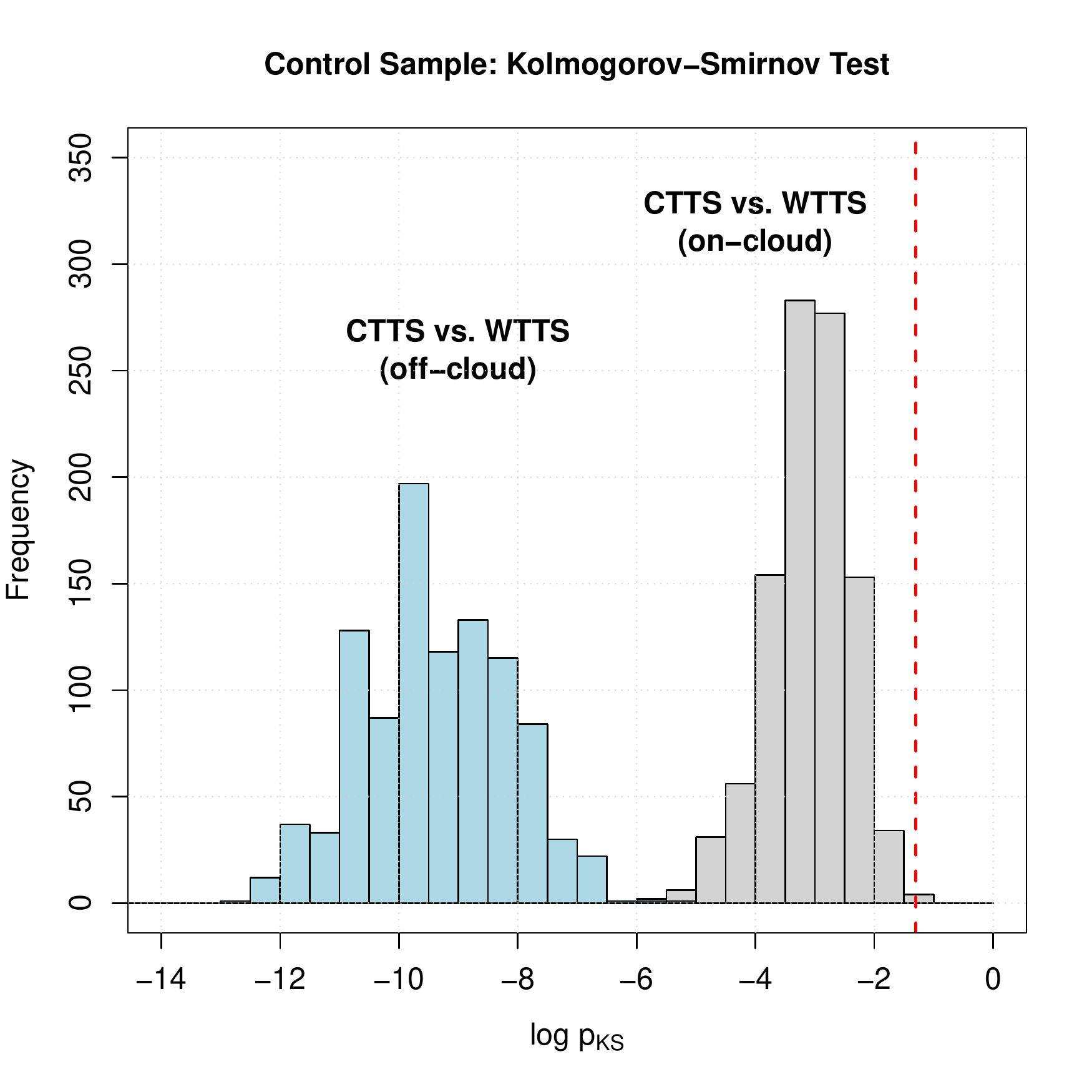}
\includegraphics[width=0.4\textwidth]{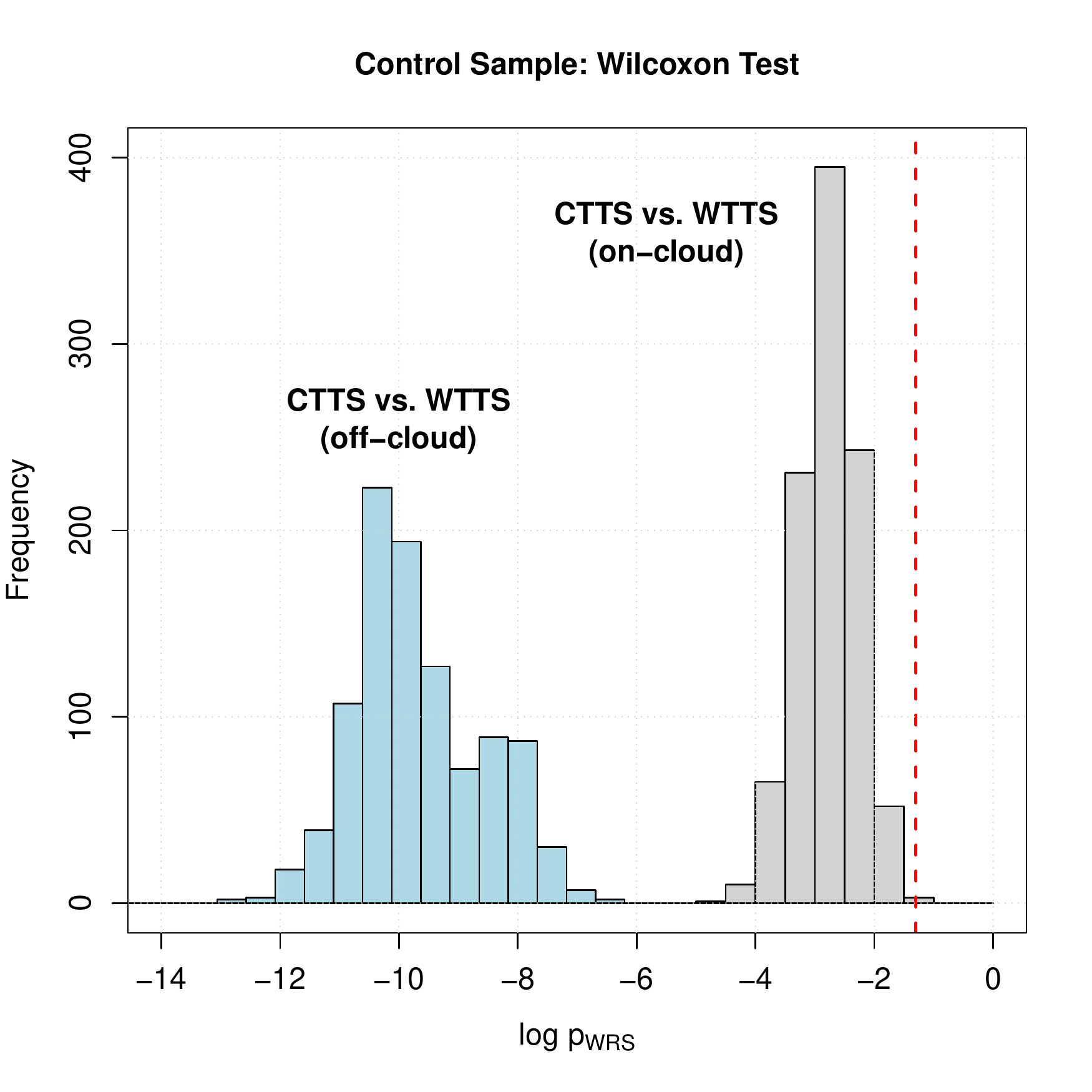}

\caption{
\label{ks_stat}
Histogram of probabilities (i.e., \textit{\emph{p-value}}) obtained from the Kolmogorov-Smirnov (\textit{left panels}) and Wilcoxon tests (\textit{right panels}) for the age distributions of the CTTSs and WTTSs in our data sample (\textit{upper panels}) and control sample (\textit{lower panels}) after 1000 Monte Carlo realizations. The red dashed line indicates the significance level of $\alpha=0.05$.  
 }
\end{center}
\end{figure*}

Furthermore, we performed Monte Carlo simulations to assess the robustness of this finding and the errors that affect our mass and age estimates. To do so, we constructed a total of 1000 synthetic samples of the Lupus association by varying, within their error bars, the effective temperatures and the remaining observables (parallaxes, magnitudes, and visual extinction) used to compute the stellar luminosities of each star (as described in Sect.~2), assuming that these variables are normally distributed. Then, we calculated the mass and age of each star from the \citet{Siess(2000)} models and performed a Kolmogorov-Smirnov and a Wilcoxon rank sum test on the mass and age distributions of the CTTSs and WTTSs for each synthetic dataset. The result of this analysis is the mass and age distributions for all synthetic realizations of the Lupus comoving stars (see Fig.~\ref{age_mass_hist}), and the histogram of probabilities that result from the statistical tests (see Fig.~\ref{ks_stat}). It is apparent from Fig.~\ref{age_mass_hist} that the \textit{\emph{on-cloud}} and \textit{\emph{off-cloud}} WTTSs define different age distributions despite exhibiting similar kinematic properties (as discussed in Paper~I). The mass distribution of both TTS subclasses shows that there is a lack of CTTSs at higher masses, while the WTTSs display a wide range of masses \citep[see also][]{Wichmann(1997a)}.

To investigate whether the different mass and age distributions of the TTSs in the Lupus association could be an artefact of the procedure used in Sect.~2 to compute the stellar luminosities, we utilized a different set of tables to convert the magnitudes and spectral types to luminosities and effective temperatures. We built a control sample by using the bolometric corrections and temperatures tabulated by \citet{Pecaut(2013)} to recompute the stellar luminosities. We then repeated the Monte Carlo simulations with our control sample as described above. The results of the KS- and Wilcoxon statistics applied to the control sample are also shown in Fig.~\ref{ks_stat}. These results confirm the different age distributions of  the CTTSs and WTTSs in the Lupus association for our data and control samples. 

To summarize, the statistical investigations reported above of the mass and age distributions of CTTSs and WTTSs contained in the Lupus association, using different conversions between observed data and stellar parameters, as well as the different evolutionary models, make us confident that these two subgroups are indeed different as far as ages are concerned. A similar conclusion was reached for the TTS population in the Taurus-Auriga association by \citet{Bertout(2007)}.

%%%%%%%%%%%%%%%%%%%%%%%%%%%%%%%%%%%%%%%%%%% 
%                                                       FIGURE 2
%%%%%%%%%%%%%%%%%%%%%%%%%%%%%%%%%%%%%%%%%%% 

%%%%%%%%%%%%%%%%%%%%%%%%%%%%%%%%%%%%%%%%%%% 
%                                                       FIGURE 3
%%%%%%%%%%%%%%%%%%%%%%%%%%%%%%%%%%%%%%%%%%% 

%%%%%%%%%%%%%%%%%%%%%%%%%%%%%%%%%%%%%%%%%%% 
%                                                       SECTION 4
%%%%%%%%%%%%%%%%%%%%%%%%%%%%%%%%%%%%%%%%%%% 

\section{T~Tauri disk lifetimes in the Lupus association}

In their study of the Taurus-Auriga SFR, \citet{Bertout(2007)} obtained a relationship between the disk lifetime and the mass of its parent star. We performed a similar study for the Lupus association, with the aim of investigating how the different environment of the Lupus SFR affects the evolution of protoplanetary disks.  We thus assumed that evolution from CTTSs to WTTSs is caused by the accretion of their circumstellar disks. Following \citet{Bertout(2007)}, the disk mass of a TTS is written as
\begin{equation}
M_{d}=\alpha(M_{\star}/M_{\odot})^{\beta}
,\end{equation}
\noindent while the mass accretion rate in units of 
$M_{\odot}/yr$ is given by  
\begin{equation}
\dot{M}_{acc}=\gamma(M_{\star}/M_{\odot})^{2.1}\, ,
\end{equation}
\noindent where $\alpha$, $\beta$, and $\gamma$ are unknowns to be determined in our analysis. This last relationship agrees with previous studies \citep[see, e.g.,][]{Muzerolle(2003),Muzerolle(2005)}, but more recent work \citep{Alcala(2014)} indicates that the exponent in Eq.~3 might be somewhat lower for low-mass CTTSs. We will come back to this point below, but for the time being we use Eq.~3 to compare our findings to those of \citet{Bertout(2007)}. The disk lifetime $\tau_{d}=M_{d}/\dot{M}_{acc}$ can then be written as
\begin{equation}
\log\tau_{d}=\log({\alpha/\gamma})+(\beta-2.1)\log(M_{\star}/M_{\odot})\, .
\end{equation}

We varied these parameters in the range of $5\leq\log(\alpha/\gamma)\leq 8$ and $-4\leq\beta\leq 4$ in steps of $0.01$ to compute $\tau_{d}$ for each star with the stellar masses given in Table~1. Then, we compared this derived disk lifetime $\tau_{d}$ for each star with its isochronal age $t_{\star}$, also listed in Table~1. The model developed by \citet{Bertout(2007)} predicts that the star is a CTTS whenever $t_{\star}\leq\tau_{d}(M_{\star})$, or a WTTS if $t_{\star}>\tau_{d}(M_{\star})$. Finally, we compared the mass distributions of the modeled CTTSs (and WTTSs) with the observed CTTSs (and WTTSs) in our data sample and ran a KS-test to find the best match between these populations for a given combination of $\alpha$, $\beta$, and $\gamma$. We adopted a probability threshold of $99.95\%$ (i.e., $p_{KS}\geq0.9995$) as in \citet{Bertout(2007)}. 
Doing so, we find $\log(\alpha/\gamma)=6.53\pm0.05$ and $\beta=2.65\pm0.10$. Our analysis is consistent with an average disk lifetime given by 
\begin{equation}
\tau_{d}\simeq3\times10^{6}\,(M_{\star}/M_{\odot})^{0.55(\pm0.10)} \rm{yr.}
\end{equation}
The position of the thus-derived disk lifetime of Lupus TTSs is plotted in the HRD of Fig.~2, together with the result obtained by \citet{Bertout(2007)} for the Taurus-Auriga SFR. 

To assess the significance of this finding, we performed Monte Carlo simulations. We applied the above strategy to the 1000 synthetic realizations of the Lupus association, as well as to the control sample  constructed in Sect.~3. The results, shown in Table~6, demonstrate that the simulations are consistent and confirm the values mentioned above. Simulations of the control sample returned lower values for the disk parameters, but these are statistically compatible with our result (within $3\sigma$). Thus, our results remain valid if we use a different set of photometric tables and spectral type to effective temperature calibration for computing the stellar luminosities and the resulting mass and age estimates. The evolution of TTSs of the Lupus association is therefore consistent with the evolutionary scenario proposed by \citet{Bertout(2007)} for the Taurus-Auriga T association.

As mentioned above, \citet{Alcala(2014)} have recently found that $\dot{M}_{acc}\propto M_{\star}^{1.8\,(\pm0.2)}$ in a sample of Lupus low-mass accreting stars. If we use this revised accretion rate estimate in Eqs.~3 and 4 and run our calculations as described above, the previous value of $\beta=2.65\pm0.10$ shifts to $\beta=2.35\pm0.10$, and the resulting exponent on the stellar mass in Eq.~5 is not affected. Similarly, if we use the mass accretion rate of $\dot{M}_{acc}\propto M_{\star}^{1.4\,(\pm0.3)}$ derived by \citet{Venuti(2014)} for a sample of CTTSs in the NGC~2264 open cluster, we find that $\beta=2.65\pm0.10$ shifts to $\beta=1.95\pm0.10$. In both cases, the different value obtained for $\beta$ affects the disk mass given by Eq.~2 in our disk model. However, the combination of the different exponent $\beta$ on the disk mass with a different mass accretion rate $\dot{M}_{acc}$ yields the same exponent on the disk lifetime in Eq.~5. Thus, our result for the average disk lifetime remains the same.

Recently, \citet{Daemgen(2013)} have investigated the impact of binary companions on the evolution of circumstellar disks and concluded that they evolve faster in binary systems. In this context, we decided to remove the binaries (and multiple systems) from our sample to inspect this point. Doing so, we recalculated the disk lifetime in Lupus and find $\log(\alpha/\gamma)=6.56\pm0.03$ and $\beta=2.69\pm0.06$. Thus, the average disk lifetime excluding binaries from the sample is $\tau_{d}\simeq4\times10^{6}\,(M_{\star}/M_{\odot})^{0.59(\pm0.06)} \rm{yr}$. This result implies that the average disk lifetime of the association is likely to be shorter when binaries (and multiple systems) are included in our analysis. However, it is important to mention that both results are still statistically compatible within their error bars. For a more direct comparison with the results obtained by \citet{Bertout(2007)} in the Taurus-Auriga region (see Sect.~5), we prefer our first solution, because the binary stars were not excluded in their study. 

%%%%%%%%%%%%%%%%%%%%%%%%%%%%%%%%%%%%%%%%%%% 
%                                                       TABLE 6
%%%%%%%%%%%%%%%%%%%%%%%%%%%%%%%%%%%%%%%%%%% 
\begin{table}[!h]
\renewcommand\thetable{6} 
\centering
\caption{
\label{tab6}
Mean and median values of the disk parameters ($\alpha$, $\beta$, and $\gamma$) obtained via Monte Carlo simulations after 1000 realizations of the data and control samples. 
}
\resizebox{9cm}{!} {
\begin{tabular}{lcccc}

\hline
&\multicolumn{2}{c}{$\log(\alpha/\gamma)$}&\multicolumn{2}{c}{$\beta$}\\
\hline
&Mean&Median&Mean&Median\\
\hline
Data sample&$6.52\pm0.05$&6.52&$2.51\pm0.10$&2.52\\
Control sample&$6.35\pm0.05$&6.36&$2.21\pm0.09$&2.22\\
\hline

\end{tabular}
}
\end{table}

Despite the success of this empirical disk model in explaining the different age distributions of the CTTSs and WTTSs in the Lupus association, we stress that the conclusions are only meaningful in a statistical sense and that our model may lead to a false classification in a few cases. Three WTTSs (RXJ1559.9-3750, RXJ1609.4-3850, and RXJ1612.0-3840) were misidentified as CTTSs in our analysis because of an inconsistency between the observed isochronal age and the computed lifetime of their disks at the $1\sigma$ level. These misidentified WTTSs are among the youngest stars in our sample. 

Another interesting point in our study is the existence of six stars (Sz91, Sz95, Sz96, Sz112, RXJ1609.9-3923, and RXJ1608.5-3847) in our sample that were first identified as CTTSs and later reclassified as transition disk objects \citep[see][]{Romero(2012),Tsukagoshi(2014)}. We have labeled them with the mention "CTTS/TD" in Table~1 to distinguish them from the remaining CTTSs. \citet{Romero(2012)} claim that RXJ1609.9-3923, the youngest transition disk object in the sample, is a \textit{``triple system with tight binary components consistent with two equally bright objects"}. Later in the text, they state that the source has an uncertain classification because its spectral energy distribution resembles a CTTS of M spectral type. Taking only the position of RXJ1609.9-3923 into
account in the HRD, our analysis suggests that this source is more likely a CTTS. The remaining transition disk objects in the sample intermingle among CTTSs and WTTSs in the HRD, and their positions are fully consistent with the disk lifetime locus that roughly separates both species (see Fig.~2). 

%%%%%%%%%%%%%%%%%%%%%%%%%%%%%%%%%%%%%%%%%%% 
%                                                       SECTION 5
%%%%%%%%%%%%%%%%%%%%%%%%%%%%%%%%%%%%%%%%%%% 
\section{Comparison with the Taurus-Auriga association}

Both the Lupus and Taurus-Auriga associations discussed in this paper have been identified by using the convergent point search method applied to proper motion data in our previous studies \citep{Bertout(2006),Galli(2013)} and the physical properties of the comoving stars were computed in the same completeness limits as described in Sect.~2. In this context, it is interesting to note that the Lupus association seems to be older than the Taurus-Auriga association. It is also apparent (see Fig.~2) that the distribution of CTTSs in Lupus is shifted toward lower masses and late spectral types, while in Taurus-Auriga they display a wider range of masses and effective temperatures \citep[see, e.g., Fig~1 of][]{Bertout(2007)}. These first results had already been suggested in previous studies by using an average distance to the Lupus SFR to compute the stellar ages (and masses) for all stars and comparing different samples of TTSs in these SFRs \citep{Hughes(1994),Wichmann(1997a)}. The individual parallaxes from Paper~I allowed us to unambiguously confirm this scenario by comparing two samples of TTSs that have been investigated with the same selection criteria.

\citet{Bertout(2007)} derived an average disk lifetime of $\tau_{d}\simeq 4\times10^{6}\,(M_{\star}/M{\odot})^{0.75}$~yr for the TTSs in the Taurus-Auriga association that contrasts with the result of $\tau_{d}\simeq3\times10^{6}\,(M_{\star}/M_{\odot})^{0.55}$~yr obtained in this paper for the Lupus association. Our heuristic disk model thus predicts different disk lifetimes for the two associations. First, the exponents defining the mass dependence are different, but the two values are still compatible with each other if one considers the uncertainties involved ($\beta=2.65\pm0.10$ in Lupus and $\beta=2.85\pm0.21$ in Taurus-Auriga). Second, regardless of the exact $\beta$ value, we note that the standard deviation values of $\log(\alpha/\gamma)$ found from the Monte Carlo simulations of Lupus and Taurus-Auriga comoving stars lead to disjointed age distributions at the 1$\sigma$ significance level for the $1M_{\odot}$ stars of both regions ($2.95-3.72$~Myr for Lupus and $3.80-4.79$~Myr for Taurus-Auriga). Although this result will need confirmation with more significant samples statistically, it gives a first indication that disk lifetimes can be different in different star-forming regions.

This is not the only indication that disk lifetimes are likely to be different in Taurus-Auriga and Lupus. For example, a comparative analysis of the HRD of each association reveals that the overlap between CTTSs and WTTSs  is more important in Lupus than in Taurus-Auriga (see Fig.~2 of this paper and Fig.~1 of \citealt{Bertout(2007)}). According to the results for the Taurus-Auriga association obtained by \citet{Bertout(2007)}, it is only the evolutionary status of the circumstellar disks that allows us to distinguish between both subgroups. In this context, the existing overlap between both species in the HRD could be indicative of a shorter disk lifetime for the TTS population in the Lupus association as illustrated in Fig.~2.

Another fact that seems to support the different disk lifetimes in each association is the age of the transition disk objects that show signs of disk dissipation and are presumably in an intermediate state between CTTSs and WTTSs with little or no detectable near-IR excess emission and significant far-IR excess \citep{Calvet(2005),Romero(2012)}. When excluding RXJ1609.9-3923 for the reasons discussed in Sect.~4, the average age of the remaining transition disks included in our sample (Sz91, Sz95, Sz96 , Sz112, and RXJ1608.5-3847) is only $1.9\pm0.3$~Myr. On the other hand, the average age of the transition disks (DM~Tau, GM~Aur, LkCa~15, and DI~Tau) in Taurus-Auriga is $5.3\pm1.4$~Myr \citep[see Table~4 of ][]{Bertout(2007)}, thus significantly older than the observed value for Lupus.

\citet{Martin(1998)} investigated the ratio of WTTSs over CTTSs in different regions of the $\rho$~Ophiuchi molecular cloud complex, and they argue that the high WTTS/CTTS ratio observed in some fields could be the result of the strong UV radiation, stellar winds, or supernova explosions from nearby massive stars that act to shorten the lifetime of circumstellar disks \citep[see also][]{Walter(1994)}. A similar argument could apply to the Lupus association, where the WTTSs greatly exceed the number of  CTTSs as compared to the Taurus-Auriga association. The Lupus cloud complex is located in an environment disturbed by the existence of hot stars from the Scorpius-Centaurus OB association and their dynamical effects, while  the low-mass SFR of Taurus-Auriga is more isolated and shows no (or weak) interaction with hot stars. In this context, the large number of WTTSs in Lupus could therefore be caused by the early disappearance of the circumstellar disks in CTTSs.

%%%%%%%%%%%%%%%%%%%%%%%%%%%%%%%%%%%%%%%%%%% 
%                                                       SECTION 6
%%%%%%%%%%%%%%%%%%%%%%%%%%%%%%%%%%%%%%%%%%% 
\section{Conclusions}

The newly derived individual parallaxes from our previous kinematic study of the Lupus association have been used in this paper to determine the photospheric luminosities and refine the masses and ages of the TTS population in this SFR. We investigated the mass and age distributions of CTTSs and WTTSs in the Lupus association and demonstrated that the probability of both subgroups being drawn from the same age distribution is very low. We also confirmed the different properties of \textit{\emph{on-cloud}} and \textit{\emph{off-cloud}} WTTSs of the Lupus association using our individual parallaxes, which is one result that had been anticipated by \citet{Wichmann(1997a)} by using an average distance estimate of the Lupus region for all stars. We concluded that the different mass and age distributions in the Lupus association can be explained by assuming that a CTTS evolves into a WTTS when the disk is accreted by the central star. Such a result had already been proposed by \citet{Bertout(2007)} for the Taurus-Auriga association, and it was now confirmed for a different SFR. 

Based on an empirical disk evolution model, we derived the average lifetime of a circumstellar disk in the Lupus association in terms of the stellar mass and compared our result with a similar study conducted for the  Taurus-Auriga association. We argued that specific properties of the Lupus association, including the more evident overlap of CTTSs and WTTSs in the HRD, the age of transition disk objects, and the ratio of WTTS/CTTS, could be the result of an earlier disappearance of the circumstellar disks in CTTSs implying that the timescale of disk dissipation in the Lupus association is shorter (than in Taurus-Auriga). 

This study represents an important step toward a better understanding of the early stages of star formation and disk evolution in different environments. It will be extended to other nearby SFRs in the future to allow for a more detailed comparative analysis of the evolution of TTSs in different associations.

%%%%%%%%%%%%%%%%%%%%%%%%%%%%%%%%%%%%%%%%%%% 
%                                               ACKNOWLEDGEMENTS
%%%%%%%%%%%%%%%%%%%%%%%%%%%%%%%%%%%%%%%%%%% 
\begin{acknowledgements}

It is a pleasure to thank Jer\^ome Bouvier and Estelle Moraux for helpful discussions that contributed to this work. We are also grateful to the referee for constructive comments that helped us to improve the manuscript. This work was funded by the S\~ao Paulo State Science Foundation (FAPESP) and the Brazilian-French cooperation agreement CAPES-COFECUB. This research made use of the SIMBAD database operated at the CDS, Strasbourg, France. This publication also made use of data products from the Two Micron All Sky Survey (2MASS), which is a joint project of the University of Massachusetts and the Infrared Processing and Analysis Center/California Institute of Technology, funded by the National Aeronautics and Space Administration and the National Science Foundation.

\end{acknowledgements}

 %%%%%%%%%%%%%%%%%%%%%%%%%%%%%%%%%%%%%%%%%%% 
%                                                       BIBLIOGRAPHY
%%%%%%%%%%%%%%%%%%%%%%%%%%%%%%%%%%%%%%%%%%% 
\bibliographystyle{aa}
\bibliography{references.bib}

\end{document}